\documentclass[useAMS,usenatbib]{mn2e}
\pdfoutput=1
\usepackage{amsmath}
\usepackage{url}
\usepackage[pdftex]{graphicx}
\title[HR\,5907: Discovery of the most rapidly rotating magnetic B-type star]{HR\,5907: Discovery of the most rapidly rotating magnetic B-type star by the MiMeS Collaboration\thanks{Based on observations obtained at the Canada-France-Hawaii Telescope (CFHT) which is operated by the National Research Council of Canada, the Institut National des Sciences de l'Univers of the Centre National de la Recherche Scientifique of France,  and the University of Hawaii.}\thanks{Based on observations collected at the European Organisation for Astronomical Research in the Southern
Hemisphere, Chile, under Prog-ID 284.D-5058.}}
\author[Grunhut et al.]{J.H. Grunhut$^{1,2}$\thanks{E-mail: Jason.Grunhut@rmc.ca}, Th. Rivinius$^3$, G.A. Wade$^2$, R.H.D. Townsend$^4$, W.L.F. Marcolino$^5$,
\newauthor
D.A. Bohlender$^6$, Th. Szeifert$^{3}$,  V. Petit$^7$, J.M. Matthews$^8$, J.F. Rowe$^{9}$,
\newauthor
A.F.J. Moffat$^{10}$, T. Kallinger$^{8,11}$, R. Kuschnig$^{8,11}$, D.B. Guenther$^{12}$, 
\newauthor
S.M. Rucinski$^{13}$, D. Sasselov$^{14}$, W.W. Weiss$^{12}$, and the MiMeS Collaboration\\
$^1$Dept. of Physics, Engineering Physics \& Astronomy, Queen’s University, Kingston, Ontario, K7L 3N6, Canada\\
$^2$Dept. of Physics, Royal Military College of Canada, P.O. Box 17000, Station Forces, Kingston, Ontario, K7K 7B4, Canada\\
$^3$ESO - European Organisation for Astronomical Research in the Southern Hemisphere, Casilla 19001, Santiago 19, Chile\\
$^4$Dept. of Astronomy, University of Wisconsin-Madison, 2535 Sterling Hall, 475 N Charter Street, Madison, WI, 53706, USA\\
$^5$Universidade Federal do Rio de Janeiro, Observat\'{o}rio do Valongo Ladeira Pedro Ant\^{o}nio, 43, CEP 20080-090, Rio de Janeiro, Brasil\\
$^6$National Research Council of Canada, Herzberg Institue of Astrophysics, 5071 West Saanich Road, Victoria, BC, V9E 2E7, Canada\\
$^7$Dept. of Geology \& Astronomy, West Chester University, West Chester, PA, 19383, USA\\
$^8$Dept. of Physics and Astronomy, University of British Columbia, 6224 Agricultural Road, Vancouver, BC, V6T 1Z1, Canada\\
$^{9}$NASA Ames Research Center, Moffett Field, CA 94035, USA\\
$^{10}$D\'{e}pt. de Physique, Universit\'{e} de Montr\'{e}al, C.P. 6128, Succursale: Centre-Ville, Montr\'{e}al, QC H3C 3J7, Canada\\
$^{11}$Institut f\"{u}r Astronomie, Universit\"{a}t Wien, T\"{u}rkenschanzstrasse 17, A-1180 Wien, Austria\\
$^{12}$Institute for Computational Astrophysics, Dept. of Astronomy and Physics, Saint Marys University, Halifax, NS, B3H 3C3, Canada\\
$^{13}$Dept. of Astronomy and Astrophysics, University of Toronto, 50 St George Street, Toronto, ON M5S 3H4, Canada\\
$^{14}$Harvard$-$Smithsonian Center for Astrophysics, 60 Garden Street, Cambridge, MA, 02138, USA\\
}

\begin{document}
\date{\today}
\pagerange{\pageref{firstpage}--\pageref{lastpage}} \pubyear{2011}
\maketitle
\label{firstpage}
\begin{abstract}
We report the discovery and analysis of a very strong magnetic field in the rapidly rotating early B-type star HR\,5907, based on observations obtained as part of the Magnetism in Massive Stars (MiMeS) project. We infer a rotation period of 0.508276$^{+0.000015}_{-0.000012}$\,d from photometric and H$\alpha$ EW measurements, making this the shortest period, non-degenerate, magnetic massive star known to date. From the comparison of {\it IUE} UV and optical spectroscopy with LTE {\sc bruce/kylie} models we find a solid-angle integrated, uniform black-body temperature of $17\,000\pm1000$\,K, a projected rotational velocity of $290\pm10$\,km\,s$^{-1}$, an equatorial radius of $3.1\pm0.2$\,$R_{\odot}$, a stellar mass of $5.5\pm0.5$\,$M_\odot$, and an inclination angle of the rotation axis to our line-of-sight of $70\pm10\degr$. Our measurements of the longitudinal magnetic field, which vary between -500 and -2000\,G, phase coherently with the rotation period and imply a surface dipole field strength of $\sim$15.7\,kG. On the other hand, from fits to mean Least-Squares Deconvolved Stokes $V$ line profiles we infer a dipole field strength of $\sim$10.4\,kG. This disagreement may result from a magnetic configuration more complex than our model, and/or from the non-uniform helium surface abundance distribution. In either case we obtain a magnetic obliquity nearly aligned with the rotation axis ($\beta=7^{+2}_{-1}\degr$). Our optical spectroscopy also shows weak variability in carbon, silicon and nitrogen lines. The emission variability in hydrogen Balmer and Paschen lines indicates the presence of a dense, highly structured magnetosphere, interpreted as a centrifugally supported, magnetically confined circumstellar disk.
\end{abstract}


\section{Introduction}
Since stars more massive than $\sim$2\,$M_\odot$ lack a significant convective envelope, which is a necessary ingredient to drive a solar-type dynamo, magnetic fields are not expected to be observed in these stars. However, the chemically peculiar, intermediate mass Ap/Bp stars have been known for over half a century to host strong, globally organised magnetic fields with surface strengths of up to tens of kG \citep[e.g. ][]{bor73}. At the hotter end of the Bp star class are the main sequence, He-strong stars \citep[e.g. ][]{boh87} that show significant enhancement and often variability in their helium lines, as is found in the archetypical star $\sigma$ Ori E \citep[e.g. ][]{land78} and the recently discovered star HR\,7355 \citep{oks10,rivi10}. In addition to the helium variability, some He-strong stars show emission variability in Balmer lines, photometric brightness variations, variable UV resonance lines, and non-thermal radio emission, most of which vary with a single period, interpreted to be the rotational period of the star \citep[e.g. ][]{ped77, walb82, shore90, leon93}. Many of these phenomena are thought to be due to the presence of a rigidly rotating, centrifugally supported magnetosphere - a region in the circumstellar environment where the stellar wind couples to the magnetic field and is forced to co-rotate with the star \citep{shore90, town05}. 

The subject of this paper, HR\,5907 (HD\,142184, V1040 Sco), is a bright ({\it V}=5.4), early type B2.5V \citep{hof91}, emission line star with a high $v\sin i$ \citep[340\,km\,s$^{-1}$; ][]{fre05}. HR\,5907 is located in the nearby Upper Scorpius OB association at a distance of $\sim$145\,pc \citep{hern05}. 

This star was originally brought to the attention of the Magnetism in Massive Stars (MiMeS) collaboration \citep{wade11} because of its $P\sim0.5$\,d Hipparcos photometric variability \citep{hub98}, which would be in good agreement with the expected rotation period for a star of this spectral type and $v_{eq}\sin i$. Additionally, archival FEROS spectra suggested that the observed H$\alpha$ profile is morphologically more similar to the profiles of other magnetic He-strong stars (such as $\sigma$~Ori~E or HR\,7355) than to classical Be stars.

In this paper we report on the discovery and first detailed investigation of the magnetic field and spectral variability properties of HR\,5907. In Sect.~\ref{obs_sec} we discuss the details of the observed spectroscopic, spectropolarimetric and photometric data, while in Sect.~\ref{ephem_sec} we re-examine the photometric and H$\alpha$ variability to refine the period. The fundamental parameters are evaluated in Sect.~\ref{fund_sect} and the spectral variability is analysed in Sect.~\ref{spec_var_sec}. In Sect.~\ref{mag_sec} we constrain the magnetic field strength and geometry using the polarimetric data. In Sect.~\ref{magneto_sec} we examine the properties of the magnetosphere and present a discussion of this study in Sect.~\ref{disc_sec}.

\section{Observations}\label{obs_sec}

\subsection{Polarimetry}

\begin{table*}
\centering
\caption{Journal of polarimetric observations listing the date, the heliocentric Julian date (2,455,000+), the number of sub-exposures and the exposure time per individual sub-exposure, the phase according to Eq.~\ref{ephemeris}, the peak signal-to-noise ratio (SNR) (per 1.8 and 3.4\,km\,s$^{-1}$ velocity bin in the ESPaDOnS spectra and FORS spectra, respectively) in the unpolarised spectra, the mean SNR per 9.0\,km\,s$^{-1}$ velocity bin in the LSD Stokes $V$ profile, and the derived longitudinal field and longitudinal field detection significance $z$ from both $V$ and $N$. All ESPaDOnS observations, except the profiles obtained on 2011-03-16, have a definite detection of a signal in Stokes $V$ (FAP $<$ 10$^{-5}$) while no detection of a signal in the diagnostic null profile is found (FAP $>$ 10$^{-3}$).}
\begin{tabular}{lccrccrrrr}
\hline
\ 	&	HJD 	& $t_{\rm exp}$ & \ &	\multicolumn{1}{c}{Peak}	& LSD & \multicolumn{2}{c}{~~~~$V$}	& 	\multicolumn{2}{c}{~~~~$N$}	\\								
\multicolumn{1}{c}{Date}	&		(2455000+)	& (s)	&  \multicolumn{1}{c}{Phase}	&	\multicolumn{1}{c}{SNR}	& SNR & $B_\ell\pm\sigma_B$ & \multicolumn{1}{c}{$z$}	&$B_\ell\pm\sigma_B$	&	\multicolumn{1}{c}{$z$}	\\
\hline
\multicolumn{9}{c}{\bf ESPaDOnS} \\
2010-02-24	&	252.0980	& $4\times350$	& 	0.8244	&      1136  &  12786	& $-1573\pm134$	&	11.8	& $-111\pm133$&	0.8	\\
2010-02-24	&	252.1174	& $4\times350$	& 	0.8626	&      1173  &  13318   & $-1201\pm128$	&	9.4	& $184\pm128$ &	1.4	\\
2010-02-24	&	252.1358	& $4\times350$	& 	0.8988	&      1285  &  14749   & $-1013\pm115$	&	8.8	& $-54\pm115$&	0.5	\\
2010-02-24	&	252.1541	& $4\times350$	& 	0.9348	&      1177  &  13519   & $-1115\pm125$	&	8.9	& $101\pm124$&	0.8	\\
2010-02-24	&	252.1730	& $4\times350$	& 	0.9719	&      1319  &  15138   & $-740\pm110$	&	6.7	& $94\pm110$&	0.9	\\
2010-02-25	&	253.0788	& $4\times350$	& 	0.7542	&      1220  &  13993   & $-1686\pm122$	&	13.9	& $-1\pm121$&	0.0	\\
2010-02-25	&	253.0972	& $4\times350$	& 	0.7902	&      1206  &  13839   & $-1733\pm124$	&	14.0	& $-141\pm123$&	1.1	\\
2010-02-25	&	253.1155	& $4\times350$	& 	0.8262	&      1259  &  14474   & $-1554\pm118$	&	13.1	& $75\pm118$ &	0.6	\\
2010-02-25	&	253.1338	& $4\times350$	& 	0.8623	&      1218  &  13998   & $-1322\pm122$	&	10.8	& $-55\pm122$&	0.5	\\
2010-02-27	&	255.1211	& $4\times350$	& 	0.7723	&      1217  &  13968   & $-1845\pm123$	&	15.0	& $-114\pm123$&	0.9	\\
2010-02-27	&	255.1395	& $4\times350$	& 	0.8084	&      1216  &  13970   & $-1663\pm123$	&	13.5	& $-167\pm123$&	1.4	\\
2010-02-27	&	255.1579	& $4\times350$	& 	0.8446	&      1234  &  14179   & $-1444\pm121$	&	11.9	& $55\pm121$&	0.5	\\
2010-02-27	&	255.1764	& $4\times350$	& 	0.8810	&      1230  &  14139   & $-1301\pm121$	&	10.8	& $35\pm120$&	0.3	\\
2010-02-28	&	256.1063	& $4\times350$	& 	0.7104	&      1265  &  14524   & $-1841\pm119$	&	15.4	& $-20\pm119$&	0.2	\\
2010-02-28	&	256.1246	& $4\times350$	& 	0.7466	&      1165  &  13372   & $-1782\pm128$	&	13.9	& $42\pm128$&	0.3	\\
2010-03-03	&	259.1164	& $4\times350$	& 	0.6328	&      1047  &  12010   & $-1709\pm150$	&	11.4	& $89\pm150$&	0.6	\\
2010-03-08	&	264.0532	& $4\times350$	& 	0.3455	&      739   &   8442   & $-1708\pm196$	&	8.7	& $-296\pm196$&	1.5	\\
2010-03-08	&	264.1429	& $4\times350$	& 	0.5221	&      1045  &  12003   & $-1821\pm155$	&	11.7	& $-31\pm155$&	0.2	\\
2010-07-23	&	400.7455	& $4\times350$	& 	0.2786	&      1245  &  13727   & $-1665\pm117$	&	14.2	& $42\pm116$&	0.4	\\
2010-07-23	&	400.8317	& $4\times350$	& 	0.4481	&      907   &  10089   & $-1708\pm176$	&	9.7	& $41\pm175$&	0.2	\\
2010-07-30	&	407.7708	& $4\times350$	& 	0.1003	&      1213  &  13394   & $-1367\pm122$	&	11.2	& $14\pm122$&	0.1	\\
2011-02-19	&	612.1578	& $4\times140$	& 	0.2183	&      748   &   8389   & $-1552\pm186$	&	8.3	& $1\pm187$&	0.0	\\
2011-02-19	&	612.1662	& $4\times140$	& 	0.2348	&      831   &   9251   & $-1656\pm171$	&	9.7	& $33\pm170$&	0.2	\\
2011-03-12	&	632.9651	& $4\times140$	& 	0.1554	&      698   &   7647   & $-980\pm386$	&	2.5	& $-3\pm387$&	0.0	\\
2011-03-12	&	632.9736	& $4\times140$	& 	0.1720	&      778   &   8554   & $-1346\pm189$	&	7.1	& $-187\pm188$&	1.0	\\
2011-03-16	&	636.9714	& $4\times140$	& 	0.0374	&      835   &   9205   & $-877\pm182$	&	4.8	& $172\pm179$&	1.0	\\
2011-03-16	&	636.9798	& $4\times140$	& 	0.0539	&      764   &   8487   & $-852\pm197$	&	4.3	& $-98\pm196$&	0.5	\\
\multicolumn{9}{c}{\bf FORS} \\                                           
2010-03-30	&	285.6680	&	$8\times1.0$	& 	0.8712	&      915   &  -  & $-750\pm251$	& 3.0	&	$441\pm266$	& 1.7 \\
2010-04-18	&	304.7476	&	$8\times1.0$	& 	0.4091	&      879   &  -  & $-2330\pm257$	& 9.1	&	$-46\pm281$	& 0.2 \\ 
2010-04-20	&	306.8728	&	$8\times1.0$	& 	0.5903	&      1272  &  -  & $-1661\pm218$	& 7.6	&	$-225\pm184$	& 1.2 \\
2010-04-25	&	311.6637	&	$8\times1.0$	& 	0.0160	&      1187  &  -  & $-1389\pm194$	& 7.2	&	$221\pm180$	& 1.2 \\
2010-04-26	&	312.7446	&	$8\times1.0$	& 	0.1426	&      1076  &  -  & $-1739\pm208$	& 8.4	&	$-207\pm207$	& 1.0 \\
2010-05-04	&	320.6415	&	$8\times1.0$	& 	0.6793	&      993   &  -  & $-2048\pm258$	& 7.9	&	$147\pm199$	& 0.7 \\
\hline
\hline
\end{tabular}
\label{obs_tab1}
\end{table*}

Between 2010 February and 2011 March, 27 high-resolution ($R\sim68\,000$) spectropolarimetric (Stokes $I$ and $V$) observations of HR\,5907 were collected with the ESPaDOnS spectropolarimeter at the Canada-France-Hawaii Telescope (CFHT) as part of the Survey Component of the MiMeS Large Program. Each spectropolarimetric observation consists of four individual sub-exposures that were processed using the {\sc upena} pipeline running {\sc libre-esprit}, following the double-ratio procedure as described by \citet{don97} to produce the Stokes $I$ and $V$ spectra. Null polarisation spectra were also produced by combining the individual four sub-exposures in such a way that the polarisation should cancel out \citep{don97}. This allows us to verify whether spurious signals are present in our reduced data.

Low-resolution ($R\sim5\,000$) spectropolarimetric observations were also collected with the FORS2 spectropolarimeter at the European Southern Observatories' (ESO) Very Large Telescope (VLT). Over six nights between February and May 2010, forty-eight exposures were collected, eight exposures per night, yielding six individual observations. The sequence of eight exposures were taken with the retarder plate positioned at angles of $-45\degr,-45\degr,+45\degr,+45\degr,-45\degr,-45\degr,+45\degr,+45\degr$ with respect to the axis of the Wollaston prism. The Stokes $V$ parameter was extracted for each observation following the method of \citet{bagn02}. A log of all our polarimetric observations is given in Table~\ref{obs_tab1}. 

\begin{figure}
\centering
\includegraphics[width=3.1in]{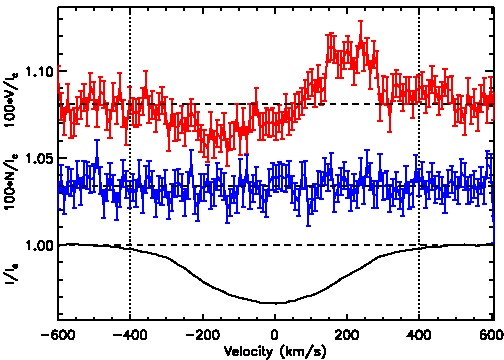}
\caption{Mean LSD Stokes $V$ (top), diagnostic null (middle) and Stokes $I$ profiles (bottom) of HR\,5907 from 2010 February 24. The $V$ and $N$ profiles are expanded by the indicated factor and shifted upwards for display purposes. A clear Zeeman signature is detected in the Stokes $V$ profiles, while the null profile shows no signal. The error bars shown in the Stokes $V$ and diagnostic null profiles represent the 1$\sigma$ uncertainties. The integration limits used to measure the longitudinal field in the ESPaDOnS LSD profiles are indicated by the dotted lines.}
\label{lsd_examp}
\end{figure}

To increase the signal-to-noise ratio (SNR) of our ESPaDOnS spectra, we applied the Least-Squares Deconvolution (LSD) procedure of \citet{don97} to all ESPaDOnS polarimetric data. As a starting point, our mask was based on a simple B2.5 star template containing only lines with intrinsic line depths greater than 10 percent of the continuum. We then proceeded to remove all lines that were blended with hydrogen lines and any that were too weak to be visible in the spectrum of HR\,5907 due to the high rotational broadening of the lines. What remained is a mask of 88 lines, most of which are helium lines. The resulting LSD profiles were computed on a spectral grid with a velocity bin of 9.0\,km\,s$^{-1}$, an example of which is shown in Fig.~\ref{lsd_examp}. The LSD method improved our sensitivity to weak Zeeman signatures by increasing the SNR by a factor of about 11 times compared to the peak SNR. We also list the improved LSD SNR measurements for each observation in Table~\ref{obs_tab1}. We find that each observation shows a clear Zeeman signature in the Stokes $V$ profile and all but the profiles obtained on 2011-03-16 have a definite detection (false alarm probability (FAP) $<10^{-5}$) according to the criteria of \citet{don02,don06}. The two profiles from this night have FAP of $1.677\times10^{-3}$ and $7.744\times10^{-3}$. In no case do we detect a significant signal in the null profile for any of our extracted mean profiles (FAP $>10^{-3}$).

The longitudinal magnetic field ($B_\ell$) and null measurements from the ESPaDOnS spectra were computed from each LSD Stokes $V$ and diagnostic null profile in a manner similar to that described by \citet{silv09}, using an integration range from -400 to 400\,km\,s$^{-1}$. We find that the $B_\ell$ varies between -740 and -1845\,G with a typical uncertainty of $\sim$120\,G. The longitudinal magnetic field was measured from the helium lines in the FORS data and were obtained in a manner similar to \citet{rivi10} and \citet[][Appendix 2]{bagn09} using a $\chi^2$ minimisation technique. The uncertainties were computed following the method outlined by \citet{rivi10} using a bootstrap Monte Carlo approach. We find a very similar range of values from -750 to -2048\,G, but with typically larger uncertainties of $\sim$250\,G. The longitudinal field measurements from each polarimetric observation are also listed in Table~\ref{obs_tab1}.

\subsection{Spectroscopy}

\begin{table*}
\centering
\caption{Journal of UVES observations listing the date, the heliocentric Julian date (2,455,000+), the exposure time, the phase according to Eq.~\ref{ephemeris}, and the peak signal-to-noise ratio (SNR) per 2.4\,km\,s$^{-1}$ velocity bin. }
\begin{tabular}{lcccclcccr}
\hline
\ 	&	HJD 	& $t_{\rm exp}$ & \ & \multicolumn{1}{c}{Peak} & \ & 	HJD 	& $t_{\rm exp}$ & \ &	\multicolumn{1}{c}{Peak}	 \\	
\multicolumn{1}{c}{Date}	&		(2455000+)	& (s)	& Phase & \multicolumn{1}{c}{SNR} & \multicolumn{1}{c}{Date}	& (2455000+)	& (s)	& Phase & \multicolumn{1}{c}{SNR}	\\
\hline
\multicolumn{5}{c}{\bf UVES 760L/760U} &                                                     \multicolumn{5}{c}{\bf UVES 437} \\                                               
2010-04-11	&	297.6922	&	12	& 	0.5280 &	348	/	284	& 2010-04-11	&	297.6923	&	12	& 	0.5281  &   444	\\
2010-04-11	&	297.6939	&	18	& 	0.5315 &      423	/	328	& 2010-04-11	&	297.6940	&	18	& 	0.5315  &   488	\\
2010-04-11	&	297.7972	&	18	& 	0.7347 &      231	/	13	& \multicolumn{1}{c}{-} 		&  - 			&        -      &      -          &  -~\, \\
2010-04-11	&	297.7994	&	18	& 	0.7389 &      446	/	329	& 2010-04-11	&	297.7995	&	18	& 	0.7391  &   452	\\
2010-04-11	&	297.9246	&	18	& 	0.9852 &      349	/	295	& 2010-04-11	&	297.9246	&	18	& 	0.9853  &   475	\\
2010-04-12	&	298.6796	&	18	& 	0.4706 &      345	/	383	& 2010-04-12	&	298.6796	&	18	& 	0.4707  &   436	\\
2010-04-12	&	298.6805	&	18	& 	0.4725 &      310	/	321	& 2010-04-12	&	298.6806	&	18	& 	0.4727  &   531	\\
2010-04-13	&	299.6495	&	18	& 	0.3789 &      511	/	318	& 2010-04-13	&	299.6495	&	18	& 	0.3789  &   384	\\
2010-04-13	&	299.6504	&	18	& 	0.3807 &      384	/	274	& 2010-04-13	&	299.6505	&	18	& 	0.3809  &   354	\\
2010-04-13	&	299.7506	&	18	& 	0.5777 &      373	/	283	& 2010-04-13	&	299.7506	&	18	& 	0.5778  &   465	\\
2010-04-13	&	299.7515	&	18	& 	0.5795 &      395	/	422	& 2010-04-13	&	299.7516	&	18	& 	0.5798  &   485	\\
2010-04-13	&	299.8095	&	18	& 	0.6937 &      447	/	312	& 2010-04-13	&	299.8096	&	18	& 	0.6938  &   440	\\
2010-04-13	&	299.8104	&	18	& 	0.6954 &      389	/	304	& 2010-04-13	&	299.8104	&	18	& 	0.6955  &   537	\\
2010-04-13	&	299.9188	&	18	& 	0.9088 &      263	/	216	& 2010-04-13	&	299.9189	&	18	& 	0.9089  &   344	\\
2010-04-13	&	299.9196	&	18	& 	0.9103 &      343	/	281	& 2010-04-13	&	299.9196	&	18	& 	0.9104  &   359	\\
2010-04-13	&	299.9204	&	18	& 	0.9119 &      269	/	209	& 2010-04-13	&	299.9204	&	18	& 	0.9119  &   335	\\
2010-04-13	&	299.9216	&	40	& 	0.9142 &      451	/	304	& 2010-04-13	&	299.9216	&	40	& 	0.9143  &   592	\\
2010-04-13	&	299.9227	&	40	& 	0.9163 &      405	/	291	& 2010-04-13	&	299.9227	&	40	& 	0.9163  &   500	\\
2010-04-13	&	299.9237	&	40	& 	0.9183 &      399	/	260	& 2010-04-13	&	299.9237	&	40	& 	0.9183  &   494	\\
2010-04-13	&	299.9248	&	40	& 	0.9205 &      507	/	310	& 2010-04-13	&	299.9248	&	40	& 	0.9206  &   491	\\
2010-04-13	&	299.9258	&	40	& 	0.9225 &      494	/	290	& 2010-04-13	&	299.9258	&	40	& 	0.9226  &   489	\\
2010-04-13	&	299.9268	&	40	& 	0.9245 &      477	/	314	& 2010-04-13	&	299.9266	&	40	& 	0.9246  &   490	\\
2010-05-30	&	346.7541	&	18	& 	0.0540 &      281	/	186	& 2010-05-30	&	346.7542	&	18	& 	0.0542  &   290	\\
2010-05-30	&	346.8314	&	18	& 	0.2061 &      221	/	183	& 2010-05-30	&	346.8314	&	18	& 	0.2062  &   271	\\
2010-05-30	&	346.8322	&	18	& 	0.2077 &      255	/	185	& 2010-05-30	&	346.8322	&	18	& 	0.2077  &   230	\\
2010-05-30	&	346.8330	&	18	& 	0.2092 &      291	/	161	& 2010-05-30	&	346.8330	&	18	& 	0.2092  &   215	\\
2010-06-08	&	356.4791	&	18	& 	0.1873 &      465	/	224	& 2010-06-08	&	356.4791	&	18	& 	0.1874  &   328	\\
\hline
\hline
\end{tabular}
\label{obs_tab2}
\end{table*}
A number of high-resolution spectra were acquired with the Ultraviolet and Visual Echelle Spectrograph (UVES) at the VLT. The instrument was used in its DIC2 437/760 setting giving us blue spectra from 375 to 498-nm and nearly continuous red spectra from 570 to 950-nm. A slit width of 0.8'' was used resulting in a resolving power of $R=50\,000$ over the entire spectrum. The details of these observations are listed in Table~\ref{obs_tab2}.

In addition to the observations already described, archival {\it IUE} UV data obtained via the NASA-MAST archive were also used to constrain the fundamental parameters of HR\,5907 (see Sect.~\ref{fund_sect}). High-resolution and low-resolution observations exist in both the SWP and LWP spectral regions, both taken in large aperture mode. The high-resolution and low-resolution spectra are in good agreement, but we have only used the low-resolution spectra (both obtained on the night of 1989-07-26) here due to the higher SNR compared to the high resolution spectra.

\subsection{Photometry}
Between 2011 April 15 and 2011 May 03 we obtained approximately uniform sampling of the photometric brightness variations of HR\,5907 using the Microvariability and Oscillations of STars (MOST) satellite. The observations were obtained with the MOST satellite in switched-target mode, meaning that observations were scheduled such that multiple targets were observed during each MOST orbital period. During our observing run, half of the orbital phase was shared with observations of Arcturus. Each observation of HR\,5907 consisted of 157 stacked 0.18\,s exposures, resulting in a total of 9846 observations.

Photometric measurements of HR\,5907 were obtained by performing aperture photometry on $20\times20$ pixel subrasters obtained from the MOST CCD photometer. We used a radius of 3 pixels for the photometric radius and a sky annulus of 8 pixels.  Centroids were measured by fitting a Gaussian Point-Spread-Function (PSF). As with all MOST photometric studies, the raw photometry shows strong photometric variations correlated to the 101 minute MOST orbital period. The variations are attributed to an increase in the measured background level of $\sim$200 ADU and to centroid shifts that map out intrapixel sensitivity changes across a CCD pixel. The photometric variations attributed to changes in the background level were measured to be $\sim$2 percent and changes in centroid position to be 0.3 percent. HR\,5907 also shows intrinsic periodic variations of $\sim$4 percent. In order to extract photometry free of instrumental effects a simultaneous fit was performed. The stray light was modelled with a third order polynomial, the intrapixel variations were modelled as a linear trend and intrinsic variations were modelled as a sinusoidal function with five additional harmonics. The fit was iterated after removing values of the centroid position and the FWHM of the PSF beyond the 90 percentile.  All data was excluded when the satellite passed through the South-Atlantic-Anomaly due to the large number of cosmic ray hits on the detector. The final reduced data set had an effective duty cycle of 26 percent with 6062 photometric points.

\begin{figure}
\centering
\includegraphics[width=3.3in]{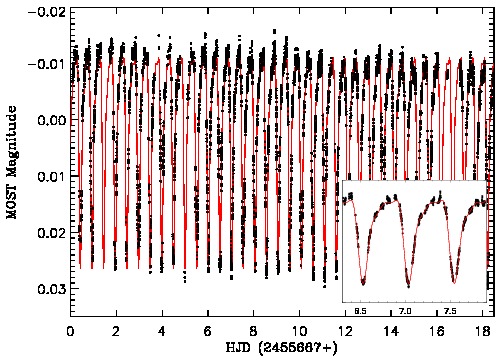}
\caption{Brightness variations of HR\,5907 as measured by the MOST satellite (black circles). The symbol size is chosen to represent the mean uncertainty of the data. The inset shows an expansion of the MOST data over a few nights to highlight the cycle to cycle variability and the clear periodicity. Also included is a red curve that corresponds to a best fit to the phased data (see Sect.~\ref{ephem_sec} for further details).}
\label{most_fig}
\end{figure}

\begin{figure*}
\centering
\includegraphics[width=6.7in]{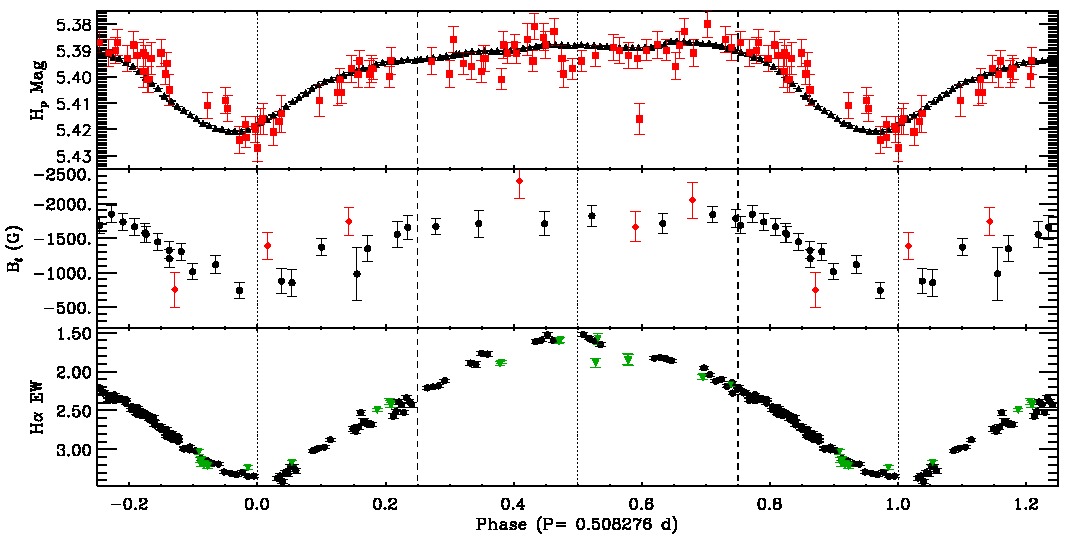}
\caption{Phased observational data according to ephemeris given in Eq.~\ref{ephemeris}: {\bf Upper panel:} MOST (black triangles) and Hipparcos (red squares) photometry. The MOST data have been binned in 0.01 phase bins and shifted to the Hipparcos magnitude for display purposes. {\bf Middle panel:} longitudinal magnetic field variations measured from ESPaDOnS (black circles) and FORS (red diamonds) data. {\bf Lower panel:} H$\alpha$ equivalent width variations measured from ESPaDOnS (black circles) and UVES (green triangles). We have also drawn dotted lines indicating phases 0.0 and 0.5, and dashed lines indicating phases 0.25 and 0.75.}
\label{phot_mag_comp}
\end{figure*}

The 18 days of observations cover roughly 36 cycles of the star, as illustrated in Fig.~\ref{most_fig}. These observations clearly demonstrate the approximately 0.5 day rotational cycle (see Sect.~\ref{ephem_sec} and Fig.~\ref{phot_mag_comp} for further details). Also shown in Fig.~\ref{most_fig} is an expansion of a few cycles that emphasises the non-sinusoidal, but periodic brightness variations. The sharpness of the photometric minimum is not consistent with brightness patches or other surface features and requires a structure that is geometrically thin, likely of circumstellar origin. We note that the observed photometric variations are consistent from cycle to cycle, but that there are additional small intrinsic variations beyond the expected scatter about the uncertainty, which cannot be attributed to any instrumental or orbital effects.

\section{Ephemeris}\label{ephem_sec}

HR\,5907 was first identified as a photometric variable star with a period of 0.508\,d based on Hipparcos data \citep{hub98}. \citet{lef09} later confirmed this variability but adopted a period that was twice as long (1.017\,d), since they assumed it was due to binary effects. We re-analysed the Hipparcos dataset using a Lomb-Scargle like technique \citep{press92} and confirm a best-fit period of $0.50831\pm0.00003$\,d. We also confirm that the current epoch of longitudinal magnetic field measurements, H$\alpha$ equivalent width (EW) variations (see Sect.~\ref{spec_var_sec}) and MOST photometry phase well with this period. However, because of the long baseline between the Hipparcos era photometry and the current MOST photometry, choosing any period within the uncertainty range can lead to a phase shift between the phased Hipparcos and MOST photometry of upwards of 0.5 cycles. Unfortunately, as the MOST data only covers about 36 rotation cycles, we cannot obtain a more precise period from this dataset. Applying the Lomb-Scargle like technique to our extensive H$\alpha$ EW dataset does provide a more precise period of $0.50825\pm0.00001$\,d, but this period is inconsistent with the Hipparcos period. In either case, we believe that the periodic variations are sufficiently non-sinusoidal that the Lomb-Scargle technique is inadequate.

We proceeded by adopting a multi-harmonic fitting technique similar to \citet{sc96}, which is ideal for non-sinusoidal periodic variations. After including contributions from the first three harmonics to both the H$\alpha$ EW variations and the Hipparcos photometry we find a best-fit period of $0.508274^{+0.000010}_{-0.000012}$\,d from the EW varations and a best-fit period of $0.508276^{+0.000015}_{-0.000012}$\,d from the Hipparcos photometry. A similar analysis of the MOST data results in a best-fit period of $0.508269^{+0.000021}_{-0.000040}$\,d period. The uncertainties on these newly derived periods are still large enough that the phasing uncertainty between the Hipparcos-era and current epoch of data is not resolved. In any event, we adopt the Hipparcos period as it provides good agreement between the Hipparcos and MOST photometry, as is expected unless significant rotational braking has occurred (see Sect.~\ref{disc_sec} for further details). Therefore, within the context of an oblique rotator model, we adopt this period as the rotational period of HR\,5907, which results in this star having the shortest known rotational period of any non-degenerate, magnetic massive star. Taking the H$\alpha$ EW maximum as $T_0$ we derive the following ephemeris:
\begin{equation}\label{ephemeris}
{\rm HJD}_{{\rm H} \alpha}^{max}=2447913.694(1)+0.508276(^{+15}_{-12})\cdot E,
\end{equation}
where the uncertainties in the last digits are indicated in brackets. Unless otherwise stated, all further data are phased according to this ephemeris.

\section{Fundamental Parameters}\label{fund_sect}
In order to investigate the fundamental parameters of HR\,5907 we utilised the third revision of the {\sc bruce} and {\sc kylie} software suite \citep[hereafter referred to as BK3; ][]{town97}. While BK3 is capable of producing theoretical spectra in both absolute flux and continuum normalised flux from either {\sc atlas9} LTE atmospheres \citep{kur92} or from NLTE {\sc tlusty} atmospheres of \citet{lanz07}, only solar abundance LTE atmospheres were used in this study. This was decided due to the fact that we found a much better overall agreement between the line strengths of C\,{\sc ii} and Si\,{\sc ii} lines in the blue region with solar abundance LTE atmospheres in comparison to solar abundance NLTE atmospheres. In our analysis, both the observed spectra and synthetic spectra were renormalised to the continuum regions surrounding the individual line profiles before the comparisons.

In the following discussion, we assume that modelling a rapidly rotating star  depends on five independent parameters, which we choose to be the equatorial velocity $v_{eq}\sin i$, the inclination angle of the rotation axis relative to the line-of-sight $i$, the effective temperature $T_{\rm eff}$, the stellar mass $M_\star$, and the stellar equatorial radius $R_{eq}$. As input, BK3 uses our estimated period $P=0.508276$\,d to constrain the equatorial values for relevant parameters. In order to determine the properties of HR\,5907, we searched through a grid of models that varied the previously listed fundamental parameters for a fixed $P$.

\subsection{Effective Temperature}
Our primary method to constrain the $T_{\rm eff}$ of HR\,5907 was from a de-reddened Spectral energy distribution (SED). We note that BK3 uses $T_{\rm eff}$ in the sense that it is the solid angle integrated, uniform black-body temperature that a star of the same surface area would need to have, such that the total luminosity is the same as the actual gravity darkened star. In this context, BK3 allows for different local temperatures over the surface of the star.

We constructed an SED for HR\,5907 by combining the {\it IUE} UV data with flux calibrated {\it UBV} photometry \citep{jasch82,khar09}, using calibrations of \citet{hynes}. To de-redden our spectra we adopted a $(B-V)_0=-0.18$ corresponding to a temperature between 17 to 18 kK. This temperature, as suggested by our initial fits to the C\,{\sc ii} 4267\,\AA\ line, is cooler than otherwise suggested by HR\,5907's B2.5 spectral classification. However, based on our initial temperature fits and the stronger than predicted helium absorption lines, HR\,5907 may in fact be a He-strong star and therefore the cooler temperature is justified. With this $(B-V)_0$ we find that $E(B-V)=0.14$, which is slightly lower than the value determined by \citet{papaj91} ($E(B-V)=0.155$) who adopted a standard B2 star template for their analysis. We do not find any significant differences in our results if we adopt this slightly higher value.

\begin{figure*}
\centering
\includegraphics[width=6.7in]{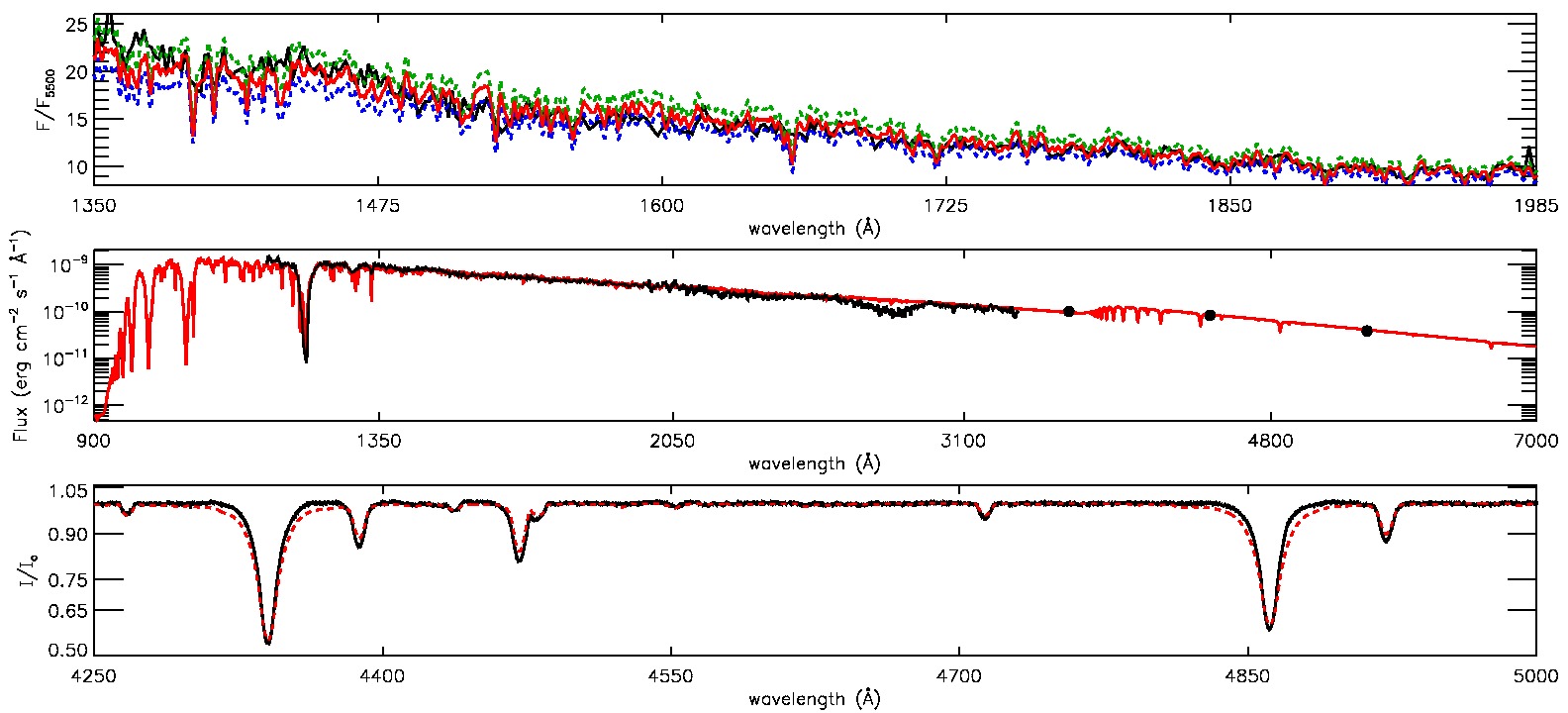}
\caption{Comparison between BK3 model and observed data: {\bf Upper panel:} {\it IUE} UV spectrum normalised to flux at 5500\,\AA\ and best-fit model with $T_{\rm eff}$ of 17000 K (red) along with models corresponding to 16000/18000\,K (blue/green). {\bf Middle panel:} {\it IUE} UV SED and Johnson {\it UBV} photometry (black) with best overall global model with $T_{\rm eff}=17000$\,K, $M_\star=5.5$\,$M_{\odot}$, $R_{eq}=3.1$\, $R_{\odot}$ and an inclination of $i=70\degr$. {\bf Bottom panel:} ESPaDOnS normalised spectrum (corresponding to observation with the least emission (24 February 2010; black)) with best-fit overall model (red). Note that the poor fit to the Balmer wings (bottom panel) is likely due to the over-normalisation of the ESPaDOnS spectrum.}
\label{spec_comp}
\end{figure*}

The SED was then extinction corrected using the parametric law of \citet{card89}, with $R_V=3.1$, and we then normalised the SED by the flux at 5500\,\AA\ so that we could constrain the temperature by fitting the slope of the UV spectrum, removing the need for any distance or luminosity corrections. As illustrated in the top panel of Fig.~\ref{spec_comp}, we find a best-fit $T_{\rm eff}=17000\pm1000$\,K when fitting the 1250-2000\,\AA\ range of the SED. We note that if we adopt an $R_V=3.9$ as found by \citet{lew09}, we find a slightly cooler temperature of 16500\,K, but still within our uncertainty. We also confirm that adopting a NLTE model atmosphere results in a best-fit $T_{\rm eff}$ within our uncertainty.

\subsection{Projected Rotational Velocity and Radius}
As previously mentioned, we could also independently constrain the $T_{\rm eff}$ by fitting the line depth of the C\,{\sc ii} 4267\,\AA\ line profile, as $i$ or $M_\star$ has very little affect on the shape of this line and it only varies slightly with $R_{eq}$. However, there is still a strong dependence of the line depth from the rotational broadening.

We simultaneously fit the $T_{\rm eff}$ and the projected rotational velocity $v_{eq}\sin i$ of the star by searching for the model that provided the best overall fit to the C\,{\sc ii} profile. We first constructed a mean profile from all of our ESPaDOnS spectra before comparing this profile to our grid of models. The ESPaDOnS profile was best fit by a model with $T_{\rm eff}=17000\pm1000$\,K and $v_{eq}\sin i=290\pm10$\,km\,s$^{-1}$. 

\begin{figure}
\centering
\includegraphics[width=3.0in]{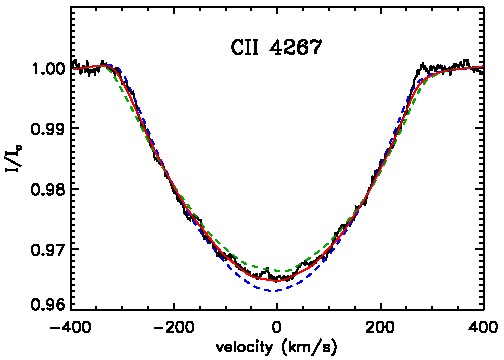}
\caption{Comparison between the mean profile from ESPaDOnS data (black) and best-fit model with $T_{\rm eff}=17000$\,K, $v\sin i=290$\,km\,s$^{-1}$ (red) and $v\sin i=280$ and $300$\,km\,s$^{-1}$ (dashed blue and green, respectively).}
\label{vsini_comp}
\end{figure}

As a check, we also compared our models with a mean profile constructed from the UVES data that resulted in a best-fit $T_{\rm eff}=17000$\,K, and a slightly lower $v_{eq}\sin i=285$\,km\,s$^{-1}$, which is still consistent with our findings from the ESPaDOnS spectra. We adopt $v_{eq}\sin i=290\pm10$\,km\,s$^{-1}$ since our mean ESPaDOnS profile is constructed from a set of observations that better span the entire rotational cycle. A comparison between our best-fit model and our mean ESPaDOnS profile is shown in Fig.~\ref{vsini_comp}. By combining our best-fit $v_{eq}\sin i$ and rotational period $P$, we find that $R_{\star, eq}\sin i\sim2.91$\,$R_{\odot}$.

\subsection{Stellar Mass and Inclination}\label{mass_inc_sect}
With $v_{eq}\sin i$ and $T_{\rm eff}$ determined, the remaining unknown model parameters are $M_\star$ and $i$, since $R_{eq}$ is constrained by $P$ and $v_{eq}\sin i$. To infer the mass of HR\,5907 we made use of the revised parallax from the Hipparcos catalogue \citep[$\pi=7.64\pm0.37$; ][]{vanle07} to obtain the absolute magnitude $M_V$ and therefore obtain HR\,5907's placement on a Hertzsprung-Russell (HR) diagram. The visual extinction $A_V$ and magnitude $V$ were taken to be 0.43 (this work) and 5.4 \citep{khar09}, respectively. Using a bolometric correction corresponding to $T_{\rm eff}$ from our model \citep[$BC=-1.6\pm0.2$; ][]{lanz07} we computed the bolometric magnitude ($M_{\rm BOL}=M_{\rm V}+BC$) and finally the luminosity $L_\star$. Our final luminosity is found to be $\log(L_\star /L_{\odot})=2.78\pm0.14$ (using the absolute bolometric corrected solar magnitude $M_\odot=4.74$), with the uncertainty found through propagation of the uncertainty in the distance and the bolometric correction.

\begin{figure}
\centering
\includegraphics[width=1.6in]{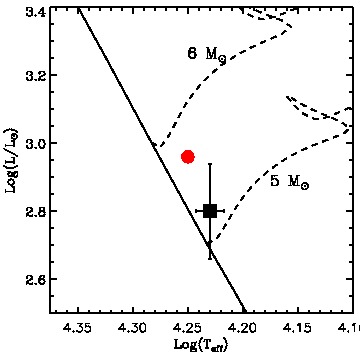}\includegraphics[width=1.6in]{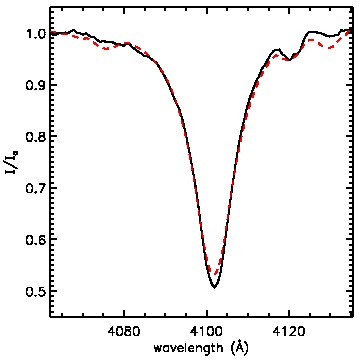}
\caption{{\bf Left panel:} Hertzsprung-Russell diagram showing the position of HR\,5907 as found in this work (black square) and as found by \citet{hern05} (red circle). Also shown are stellar evolutionary tracks (dashed) and the zero-age main sequence (solid) from \citet{bert09}. {\bf Right panel:} Comparison of our best-fit model ($T_{\rm eff}=17$\,kK, $M_\star=5.5$\,$M_\odot$, $R_{eq}=3.1$\,$R_\odot$) with the FORS observation from 25 April 2011 showing the least emission, for the H$\delta$ region.}
\label{hr_diag}
\end{figure}

As shown in Fig.~\ref{hr_diag}, we compared HR\,5907's placement with available Padova evolutionary tracks \citep[{\it Y}=0.26; ][]{bert09} and find that HR\,5907 has a mass of $\sim5.2$\,$M_{\odot}$. Also included in Fig.~\ref{hr_diag} are the results of \citep{hern05} who find HR\,5907 slightly hotter and more luminous, but still consistent with our findings. The Padova evolutionary tracks may lack any effects due to rotation, but according to the work of \citet{maeder10}, the inclusion of rotation would shift the evolutionary track such that a star of a particular mass and age would be slightly cooler and slightly brighter when rotation is included. Therefore, if we used tracks that included rotation, then based on HR\,5907's position we would only expect to find small changes (within our uncertainty) that would result in this star being slightly closer to the main sequence and having a slightly higher mass.

We can also constrain the mass and inclination from fits to the Balmer wings, which are sensitive to surface gravity ($g=G\,M_\star/R^2_\star$). Unfortunately, this can be quite problematic with HR\,5907 as there is considerable emission in the wings of the Balmer lines due to circumstellar plasma (see Sect.~\ref{magneto_sec}). However, we still proceeded by using the FORS spectrum from 25 April 2011 that showed the least emission in the Balmer wings. The long-slit FORS spectra are ideal since they are less susceptible to the over-normalisation of the Balmer lines that can occur with the multi-order ESPaDOnS and UVES spectra. We proceeded by searching for the model (at a fixed $T_{\rm eff}$, but varying mass and radius) that provided the best fit to the wings of the H$\delta$ line profile, since lower Balmer series lines are more affected by the circumstellar plasma, and higher Balmer series lines were more difficult to ensure consistent normalisation between the model and observed spectrum. An overall best-fit model was found with $M_\star=5.5\pm0.5$\,$M_{\odot}$ and $R_{eq}=3.1\pm0.1$\,$R_{\odot}$ (a comparison is shown in the right panel of Fig.~\ref{hr_diag}), consistent with the mass estimates derived from HR\,5907's placement on the HR diagram. If we adopt $M_\star=5.5\pm0.5$\,$M_{\odot}$ and $R_{eq}=3.1\pm0.1$\,$R_{\odot}$ this implies that values of $i=70^{+20}_{-10}\degr$ are allowed.

Independently, we can also use the BK3 models to fit the absolute, extinction-corrected SED of HR\,5907. Using the previously stated parallax value, we first corrected the SED to an absolute distance consistent with the BK3 models. We then proceeded by fitting the {\it IUE} UV spectra (between 1250-2000\,\AA) with the BK3 models. From this method we also find a best-fit $R_{eq}=3.1\pm0.1$\,$R_{\odot}$ and $i=70\pm10\degr$ for masses around 5\,$M_\odot$, as shown in the middle panel of Fig.~\ref{spec_comp}. However, if we use the $R_V$ value of \citet{lew09} and the slightly cooler temperature that results from those fits, we would find a slightly smaller equatorial radius of 2.9\,$R_{\odot}$. We therefore adopt a slightly higher uncertainty of 0.2\,$R_{\odot}$.  A summary of the fundamental stellar parameters is listed in Sect.~\ref{disc_sec}. A comparison between our best-fit model and the ESPaDOnS spectrum with the lowest emission is included in the bottom panel of Fig.~\ref{spec_comp}.

\section{Line Variability}\label{spec_var_sec}
Because of HR\,5907's rapid rotation and the need to sample fast photospheric variations, we opted to use the individual sub-exposures that make up the polarimetric ESPaDOnS observations for all our line profile variability analysis. The individual sub-exposures are still of sufficient quality, with SNRs comparable to the UVES data. 

\begin{figure*}
\centering
\includegraphics[width=6.7in]{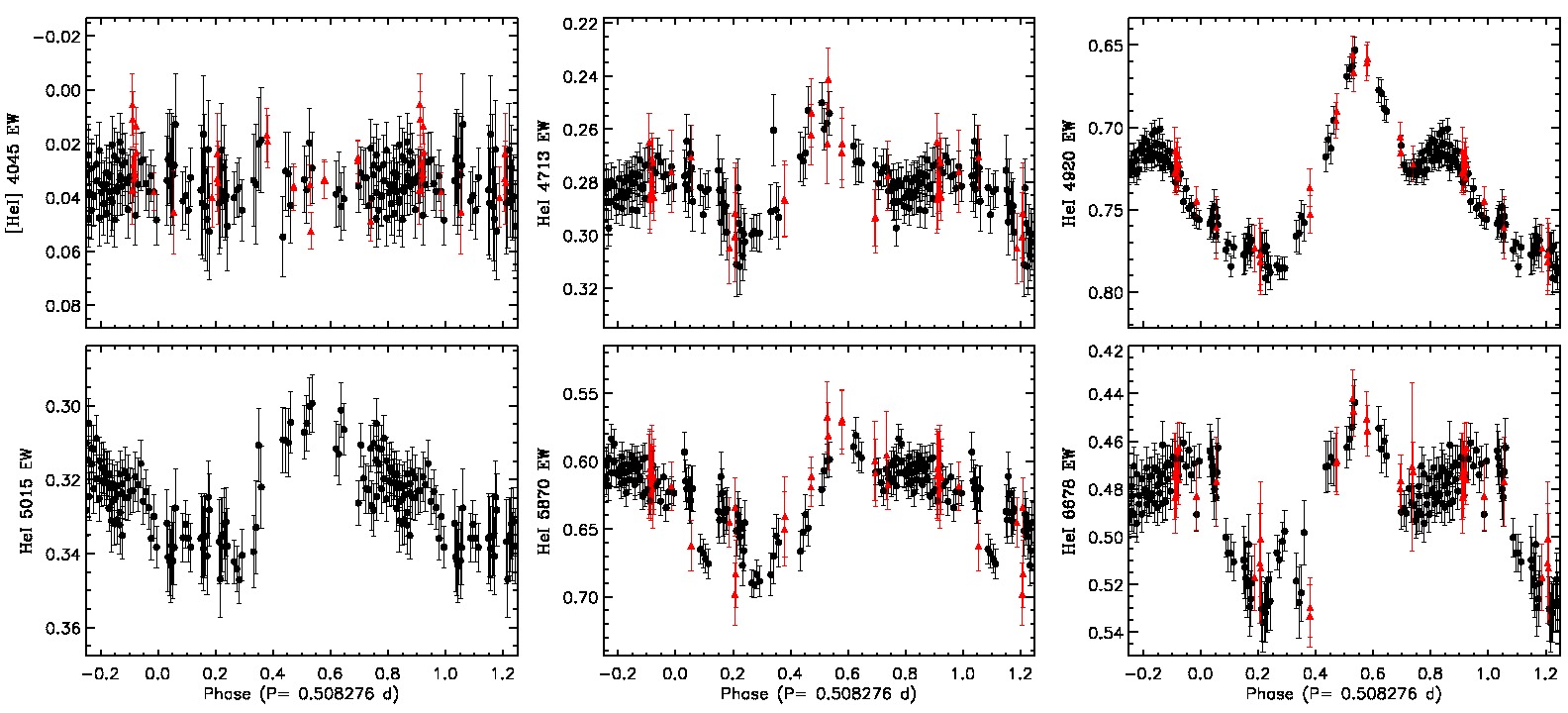}
\caption{Phased equivalent width variations of selected helium lines in the spectrum of HR\,5907, measured from ESPaDOnS (black circles) and UVES (red triangles) data.}
\label{he1_ew}
\end{figure*}

\begin{figure*}
\centering
\includegraphics[width=6.7in]{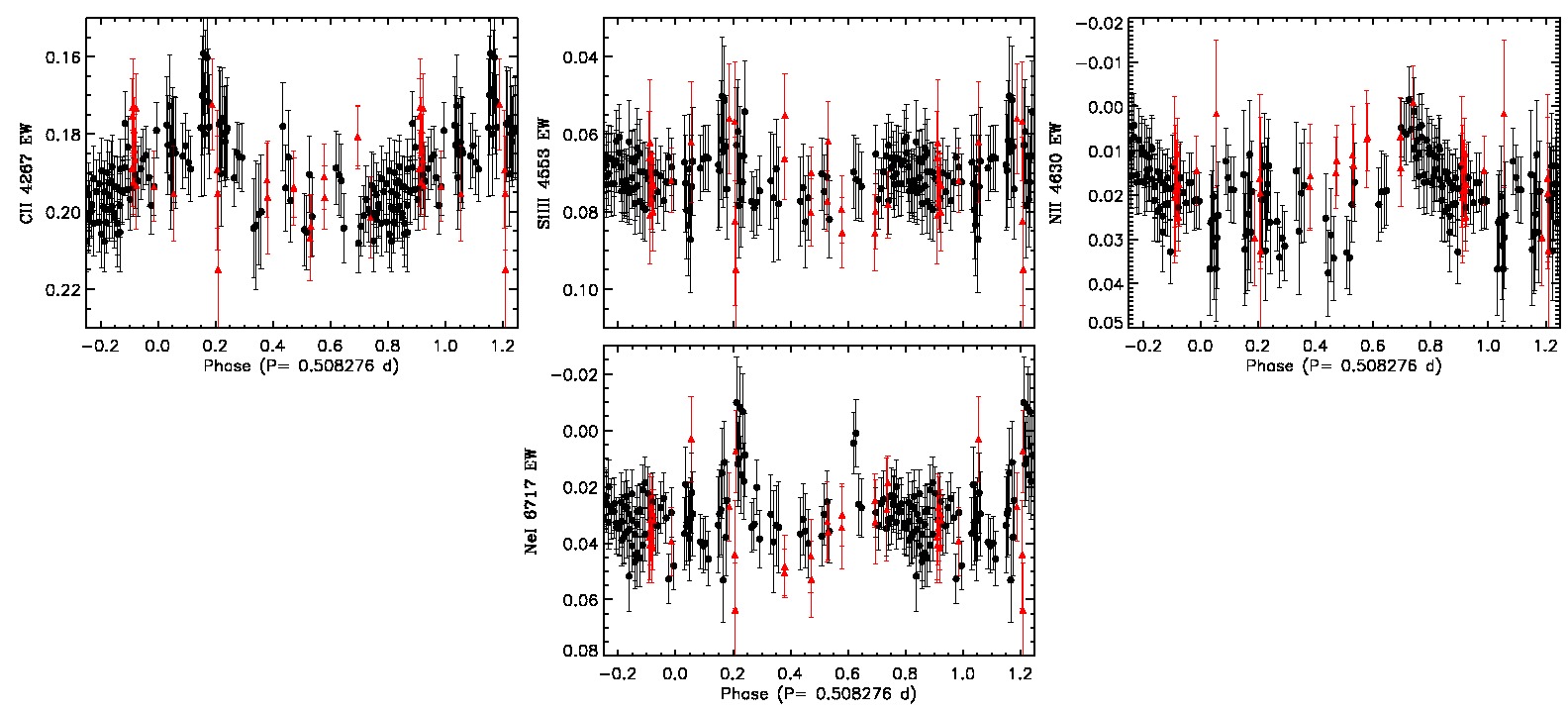}
\caption{Phased equivalent width variations of selected metal lines in the spectrum of HR\,5907, measured from ESPaDOnS (black) and UVES (red) data.}
\label{other_ew}
\end{figure*}

To begin, we can characterise the variability from variations in the equivalent widths (EWs) of the spectral lines. Before measuring the EW,
each spectral line was re-normalised to the surrounding continuum, and a telluric correction algorithm was applied to regions redward of 5790\,\AA. The EWs were then obtained by numerically integrating over the line profile. The 1$\sigma$ uncertainties were computed by adding the individual pixel uncertainties in quadrature. For the UVES data, a single uncertainty value was assigned to each pixel, determined from the RMS scatter in the continuum regions surrounding the line profile. The EW variations for a number of spectral lines were measured and then phased  with the ephemeris given in Eq.~\ref{ephemeris}, with the results presented in Figs.~\ref{phot_mag_comp}, \ref{he1_ew} and \ref{other_ew}. We find small systematic offsets between the UVES data and the ESPaDOnS data on the order 0.01\,\AA\ for most helium and metallic lines. However, a larger offset is found in the broader Balmer and Pashen lines. We do not find any significant differences in the morphology of the line profiles between the two datasets at the same phase, but we do find weak but wide-scale systematic differences. We attribute these systematic offsets to inconsistencies in the normalisation between the two datasets, which is common due to complexity of extraction and normalisation of echelle spectra. Therefore, the UVES EW values have all been corrected to best match the ESPaDOnS measurements by subtracting fixed offsets between the two datasets.

As depicted in Fig.~\ref{he1_ew}, helium lines show obvious EW variations, consistent with the rotational period, as measured from our high-resolution spectra. The observed EW variations do not vary in a similar manner to the circumstellar H$\alpha$ variations depicted in Fig.~\ref{phot_mag_comp} and we speculate that they are due to photospheric abundance patches, consistent with findings from investigations of other He-strong/variable stars \citep[e.g. ][]{boh88}.

The clearest variability is seen in the He\,{\sc i} 4920\,\AA\ line, which reaches maximum absorption (corresponding to a minimum EW in Fig.~\ref{he1_ew}) around phase 0.25 and minimum absorption around phase 0.55. Another local minimum is found around phase 0.85, indicating a relatively complex helium abundance pattern in the photosphere of HR\,5907. In contrast to the large EW variations seen in He\,{\sc i} 4920\,\AA, very little variability is observed in the forbidden [He\,{\sc i}] 4045\,\AA\ line. In the other helium lines shown in Fig.~\ref{he1_ew}, we see a variability pattern similar to the 4920\,\AA\ line, but with a smaller amplitude. In comparison to other He-strong stars such as $\sigma$~Ori~E \citep[e.g. ][]{land78, rein00, smith07} or HR\,7355 \citep{rivi10} we find that HR\,5907 shows relatively weak variability in its helium lines, of $\sim$20 percent in EW. 

We also investigated the EW variability of a number of other photospheric absorption lines, the results of which are displayed in Fig.~\ref{other_ew}. In comparison to the helium lines, very little variability is seen in these lines. Only the lines of carbon and nitrogen, as evidenced in Fig.~\ref{other_ew} (top left and top right panels) show any obvious signs of periodic variations. The EW curve for carbon is approximately sinusoidal, with maximum absorption around phase 0.2 and minimum absorption at about phase 0.7. The nitrogen EW curve is also simple, but appears to be a non-sinusoidal variation.

\begin{figure*}
\centering
\includegraphics[width=2.19in]{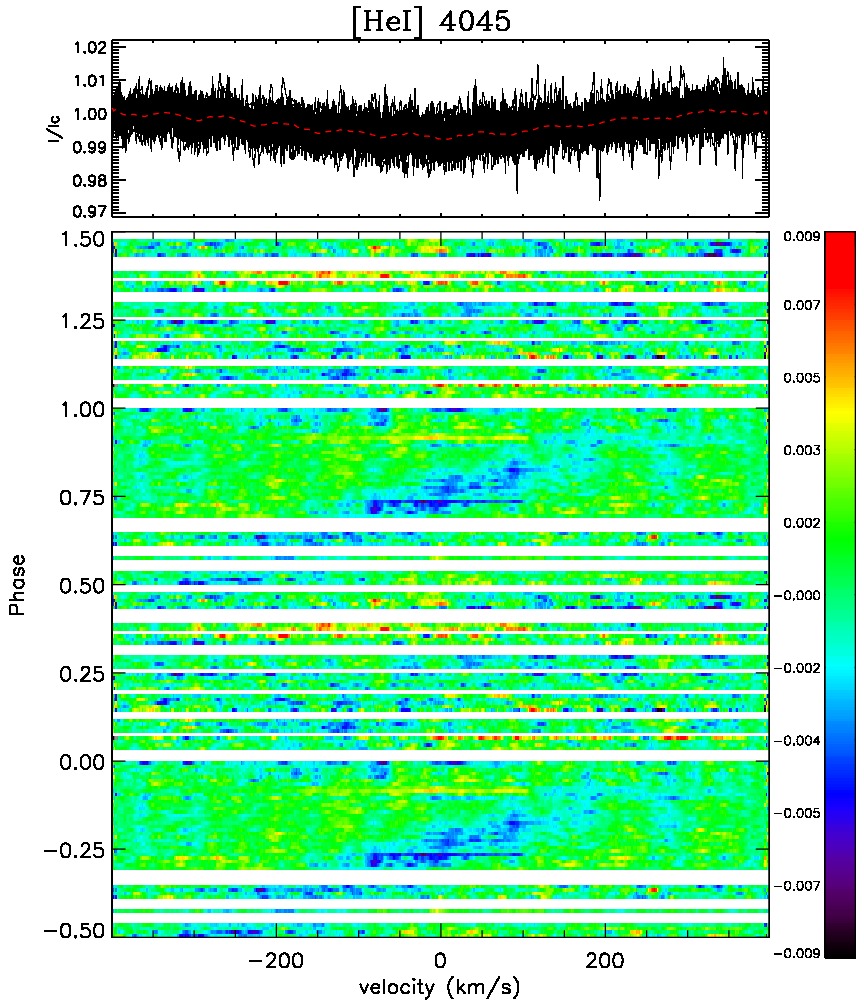}
\includegraphics[width=2.19in]{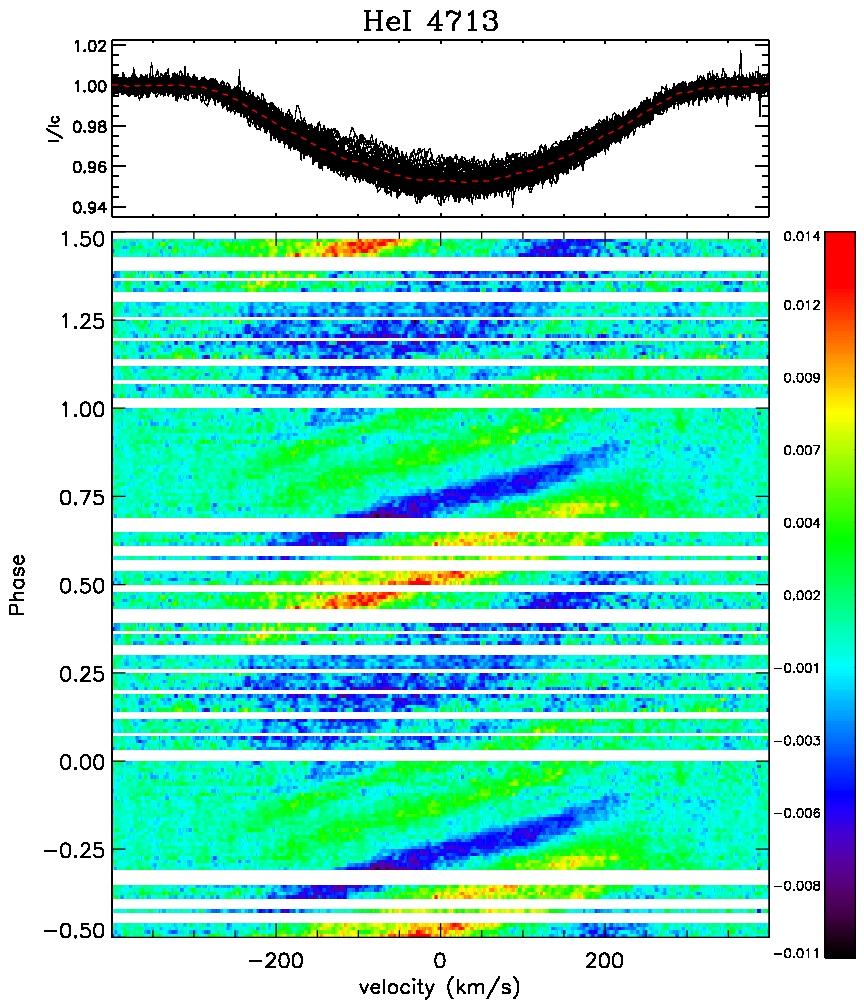}
\includegraphics[width=2.19in]{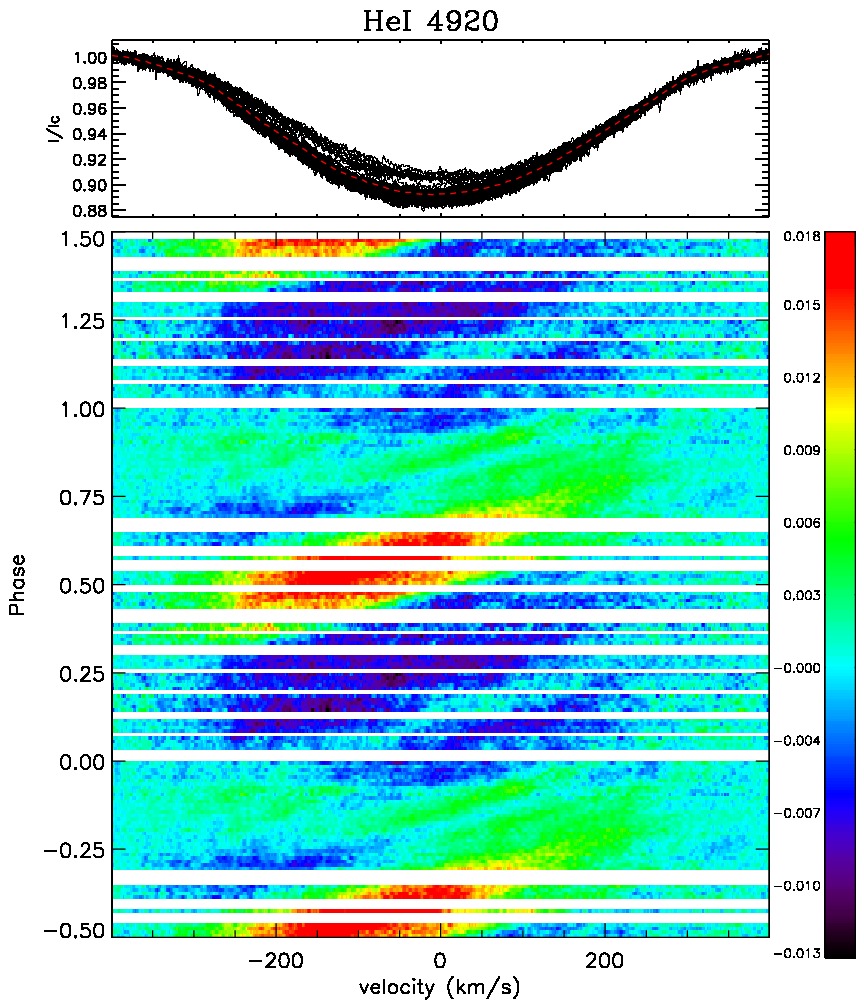}
\\
\includegraphics[width=2.19in]{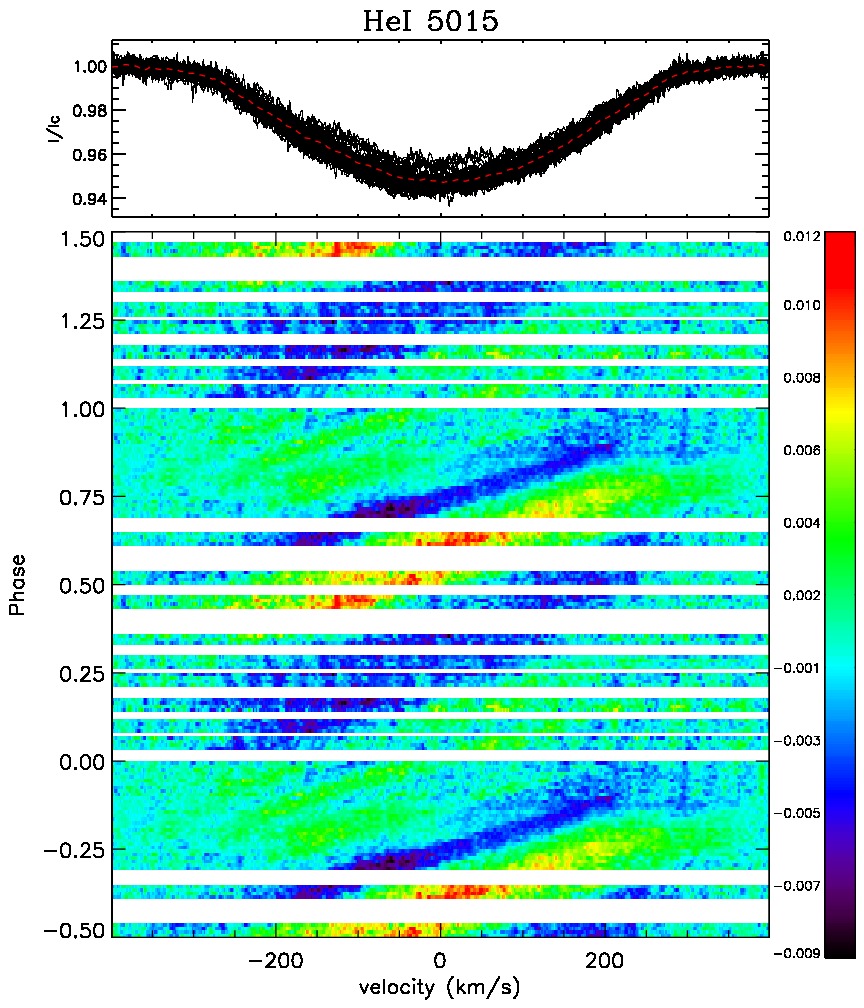}
\includegraphics[width=2.19in]{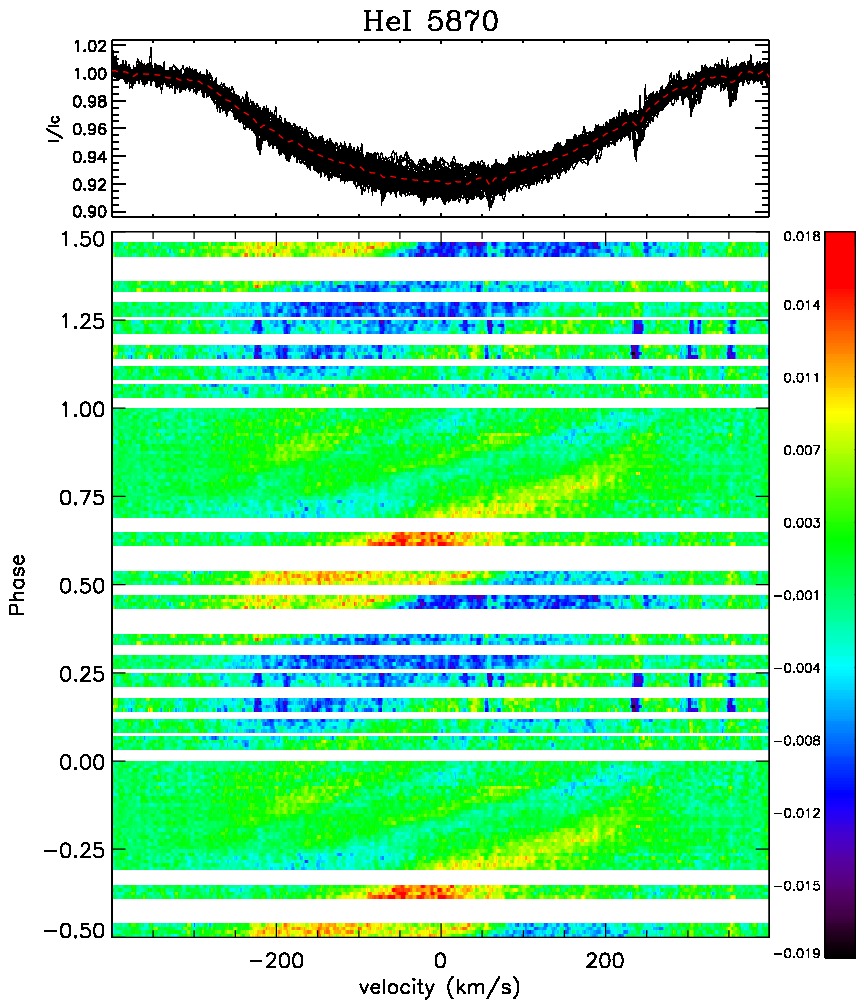}
\includegraphics[width=2.19in]{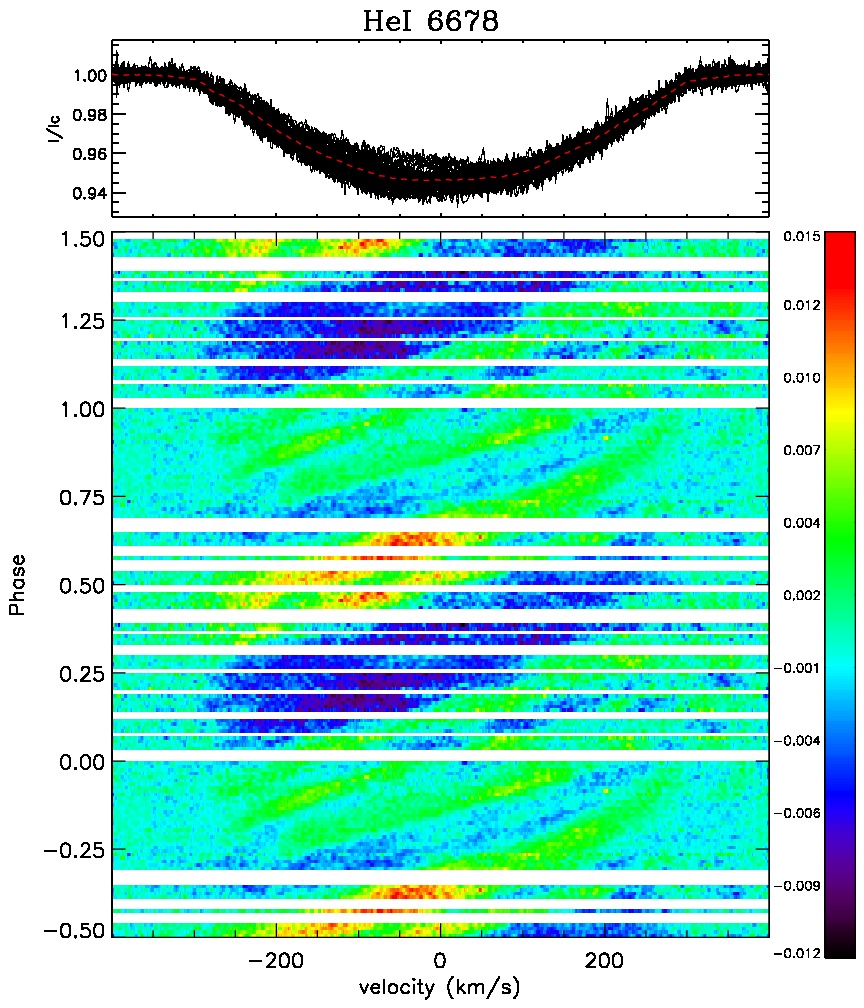}
\caption{Phased variations of photospheric helium lines.}
\label{he_dyn}
\end{figure*}

\begin{figure*}
\centering
\includegraphics[width=2.19in]{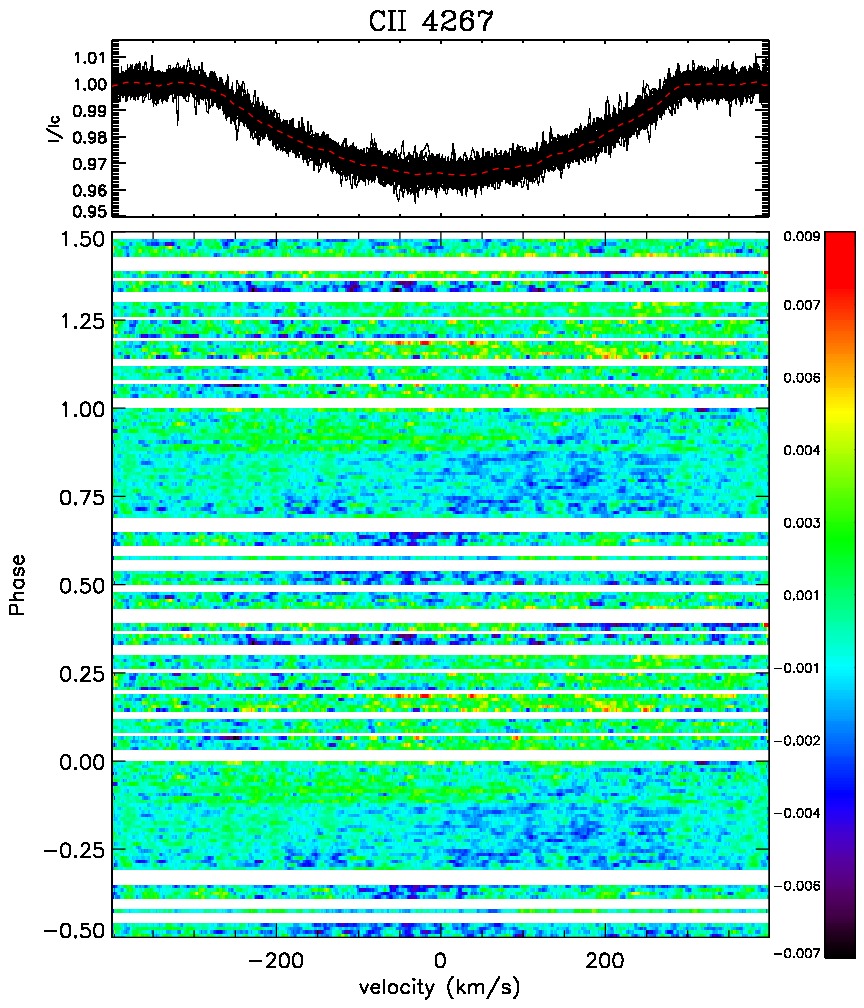}
\includegraphics[width=2.19in]{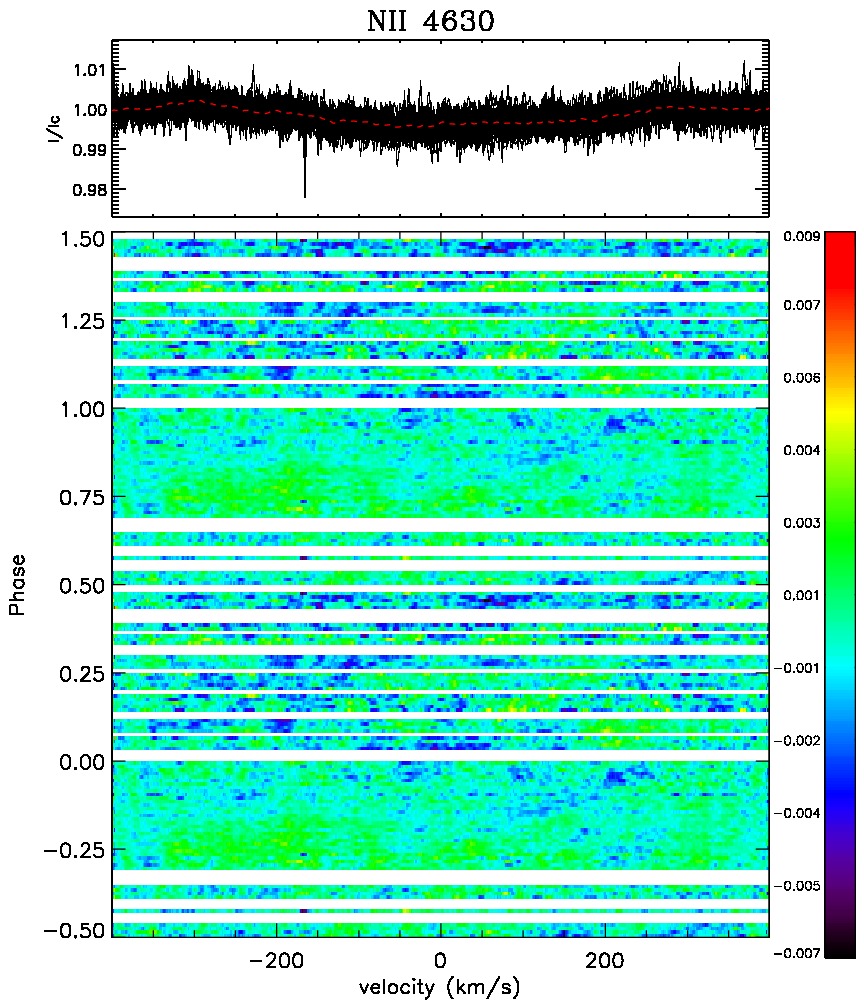}
\includegraphics[width=2.19in]{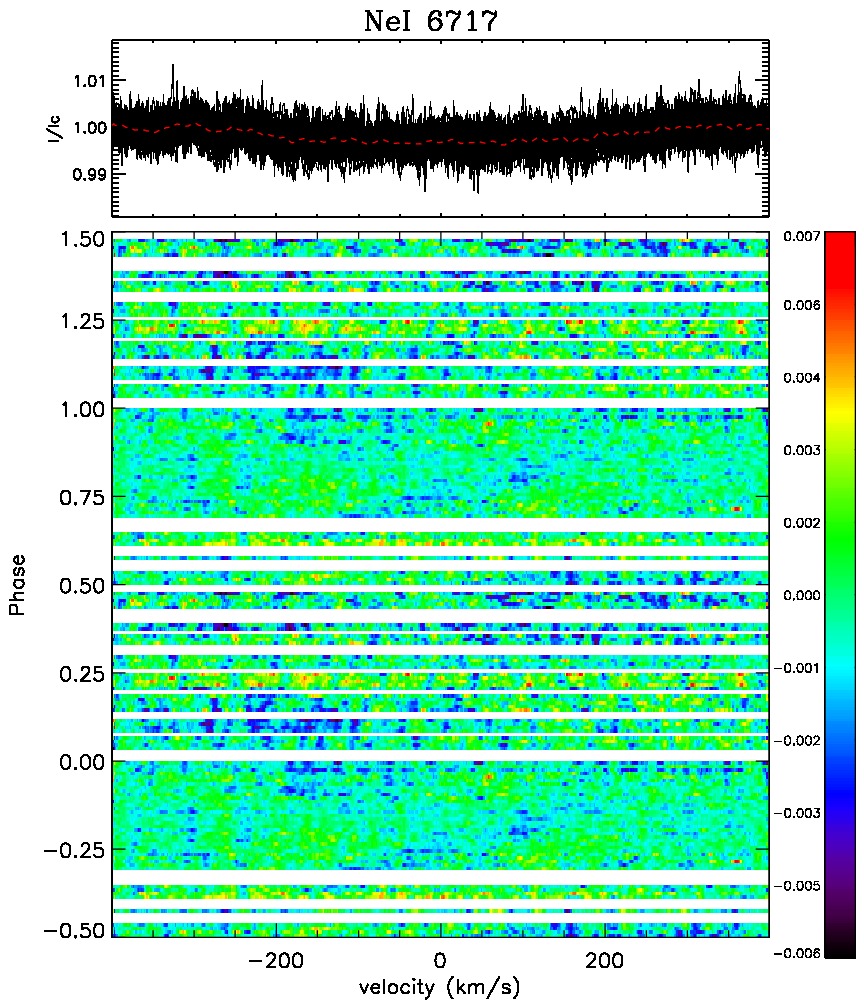}
\\
\includegraphics[width=2.19in]{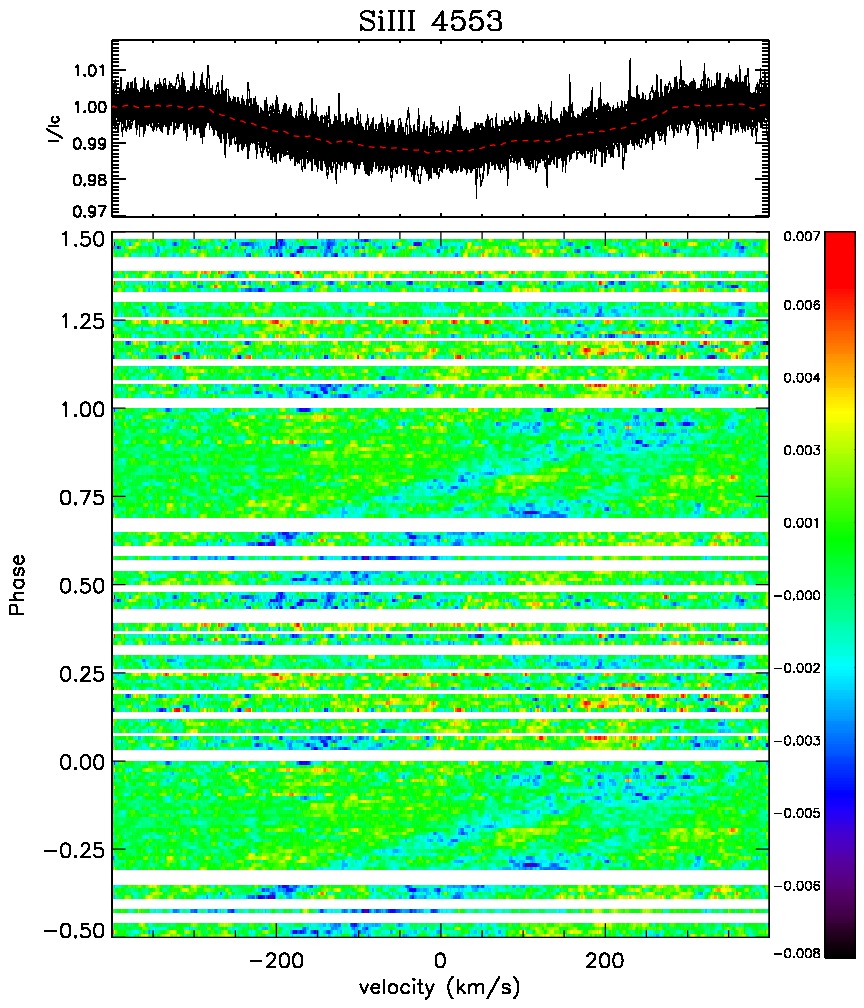}
\caption{Phased variations of selected metallic photospheric lines.}
\label{other_dyn}
\end{figure*}

We can further investigate the photospheric variations by examining the phased intensity variations of these line profiles, commonly referred to as ``dynamic spectra", as shown in Figs.~\ref{he_dyn} and \ref{other_dyn}. In these figures, individual spectra are represented as horizontal bands with intensity encoded by different colours and are stacked vertically according to their phase to illustrate time variability. To further highlight these variations, we have also subtracted from each profile the mean spectrum formed by averaging all the profiles. 

The dynamic spectrum of the [He\,{\sc i}] 4045\,\AA\ line (Fig.~\ref{he_dyn} upper left panel) shows a very weak but simple pattern with one clear enhanced absorption feature travelling from negative to positive velocities, crossing $v_{sys}=0$\,km\,s$^{-1}$ at about phase 0.75 (where $v_{sys}$ represents the systemic velocity of HR\,5907). The absorption feature is likely a region of the stellar photosphere where helium is found to be over-abundant compared to the mean helium surface distribution. This feature is also present in all the other helium lines shown in Fig.~\ref{he_dyn}, but appears to be the less prominent absorption feature in the other helium lines. A significantly larger and more absorptive feature is found to cross  $v_{sys}=0$\,km\,s$^{-1}$ at about phase 0.25 in all the other helium lines. Additionally, these stronger helium lines also show a clearly more complex dynamic spectrum. In these lines there also appears to be the presence of a strong pseudo-emission feature (this feature only appears in emission relative to the mean profile and represents a region of the stellar photosphere where helium is under-abundant with respect to the mean helium surface distribution), which crosses $v_{sys}=0$\,km\,s$^{-1}$ around phase 0.6. In some of the lines (e.g. He\,{\sc i} 4713, 5015, and 6678\,\AA) additional pseudo-emission features can been found with one crossing $v_{sys}=0$\,km\,s$^{-1}$ around phase 0.8 and the other around phase 0.0. These two features are much narrower than the feature that crosses at phase 0.6, and do not reach the same intensity of pseudo-emission. 

If we look at the dynamic spectra of the metallic lines (Fig.~\ref{other_dyn}), we find a single extremely weak pseudo-emission feature that crosses $v_{sys}=0$\,km\,s$^{-1}$ around phase 0.2 in C\,{\sc ii} or an enhanced absorption feature crossing $v_{sys}=0$\,km\,s$^{-1}$ around phase 0.6 (upper left panel). We have also tentatively identified a weak absorption feature in N\,{\sc ii} and Ne\,{\sc i} crossing $v_{sys}=0$\,km\,s$^{-1}$ around phase 0.2, which appears to be in anti-phase with the absorption feature found in the C\,{\sc ii} line. However, these lines are very weak, making it difficult to distinguish any features relative to the noise level. In the dynamic spectrum of Si\,{\sc iii} 4553\,\AA\ (bottom panel), we find two absorption features closely spaced in phase, with one crossing $v_{sys}=0$\,km\,s$^{-1}$ around phase 0.65 and the other crossing around phase 0.8. These variations appear to be in phase with the C\,{\sc ii} variations.

\section{Magnetic Field Geometry}\label{mag_sec}
To investigate the magnetic field geometry of HR\,5907 we assume that the field can be described by the dipole Oblique Rotator Model (ORM). This model is characterised by four parameters: the phase of closest approach of the magnetic pole to the line of sight $\phi_0$, the inclination of the stellar rotation axis $i$, the obliquity angle between the magnetic axis and the rotation axis $\beta$ and the dipole polar strength $B_d$. Our first approach was to model the longitudinal field curve, as shown in Fig.~\ref{phot_mag_comp}. This was carried out using a $\chi^2$ minimisation to compare the observed longitudinal field curve to a grid of computed longitudinal field curves to determine $B_d$ and $\beta$. For this analysis we have assumed $i=70\degr$ (as determined in Sect.~\ref{mass_inc_sect}) and a limb darkening coefficient of 0.4.

\begin{figure}
\centering
\includegraphics[width=2.7in]{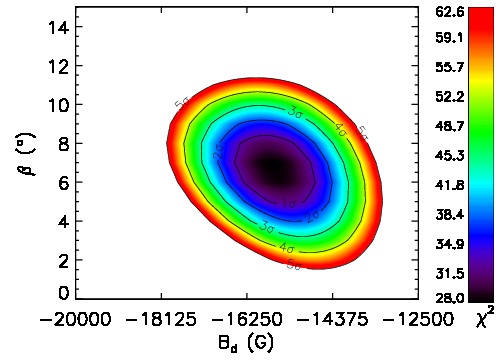}
\caption{$\chi^2$ landscape of dipole field strength $B_d$ versus magnetic obliquity $\beta$ permitted by the longitudinal field variation of HR\,5907, assuming $i=70\degr$.}
\label{chisq_land}
\end{figure}

\begin{figure}
\centering
\includegraphics[width=3.2in]{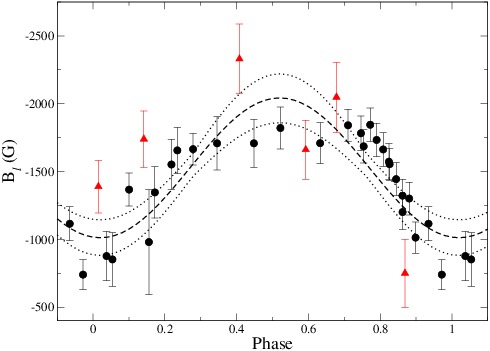}
\caption{Longitudinal field variations of HR\,5907 from ESPaDOnS (black circles) and FORS (red triangles) compared to our best-fit model curve (dashed) with $B_d=15700$\,G and $\beta=7\degr$. The dotted lines indicated the 1$\sigma$ limits permitted by our fits.}
\label{long_curve}
\end{figure}

The resulting $\chi^2$ landscape from our fits is shown in Fig.~\ref{chisq_land}. Our results indicate that we are viewing the negative (southern) hemisphere of the magnetic field and that HR\,5907 hosts a field with a polar surface strength of $B_d=15700^{+800}_{-900}$\,G, an obliquity angle of $\beta=7^{+1}_{-2}\degr$ and that $\phi_0=0.48$. Despite relatively small formal uncertainties our best-fit model is not capable of fitting the measured longitudinal field measurements at each phase, as illustrated in Fig.~\ref{long_curve}, and as indicated by the best-fit reduced $\chi^2=2.0$ from the FORS and ESPaDOnS datasets combined, or 2.1 from ESPaDOnS data alone. 

\begin{figure*}
\centering
\includegraphics[width=1.65in]{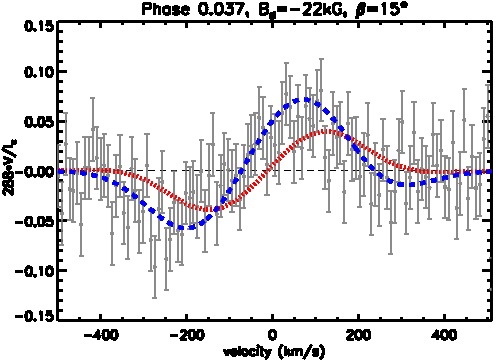}
\includegraphics[width=1.65in]{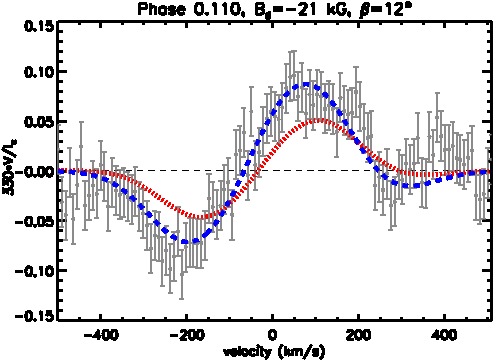}
\includegraphics[width=1.65in]{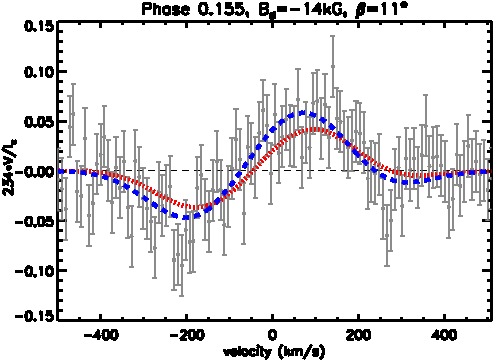}
\includegraphics[width=1.65in]{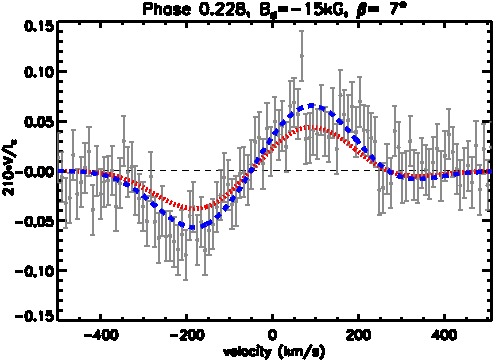}
\\
\includegraphics[width=1.65in]{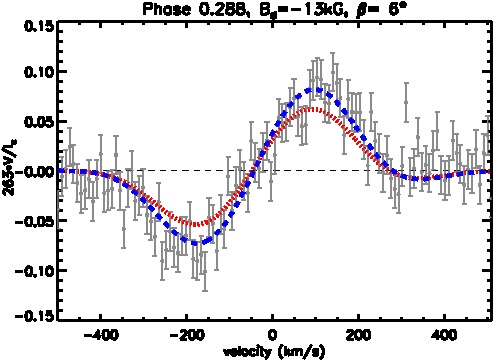}
\includegraphics[width=1.65in]{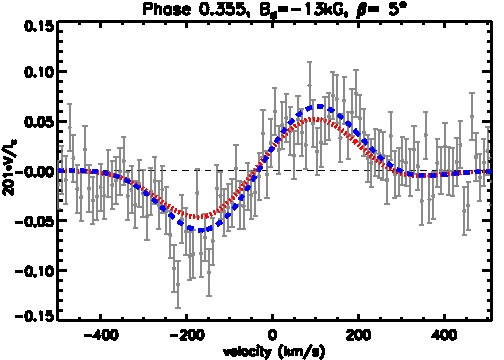}
\includegraphics[width=1.65in]{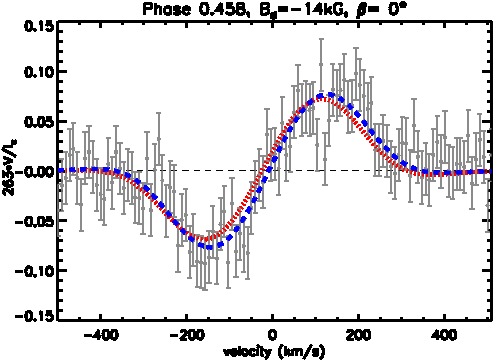}
\includegraphics[width=1.65in]{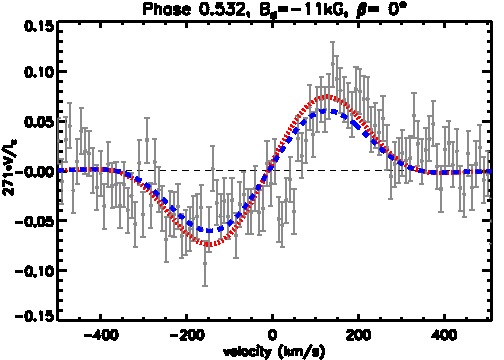}
\\
\includegraphics[width=1.65in]{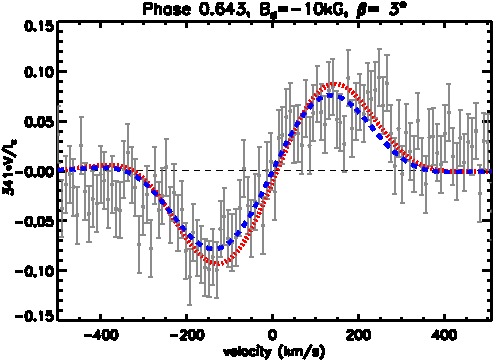}
\includegraphics[width=1.65in]{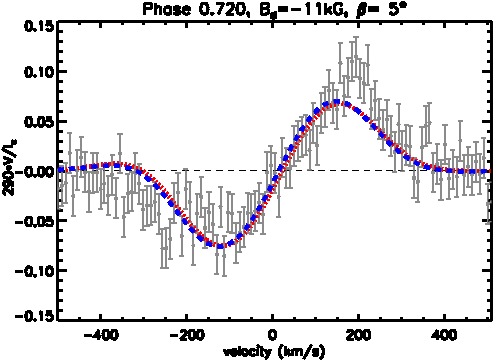}
\includegraphics[width=1.65in]{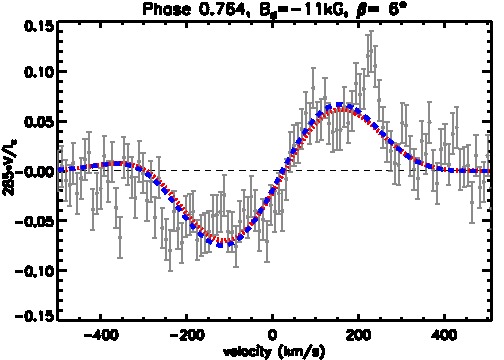}
\includegraphics[width=1.65in]{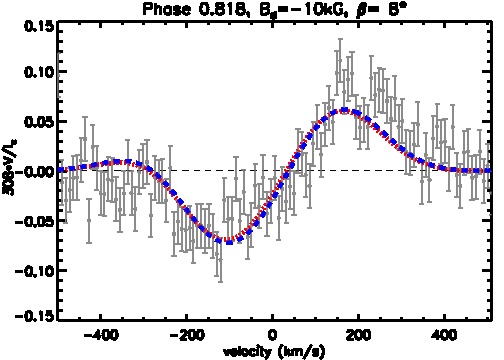}
\\
\includegraphics[width=1.65in]{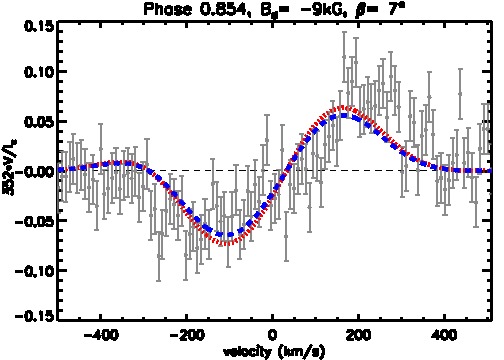}
\includegraphics[width=1.65in]{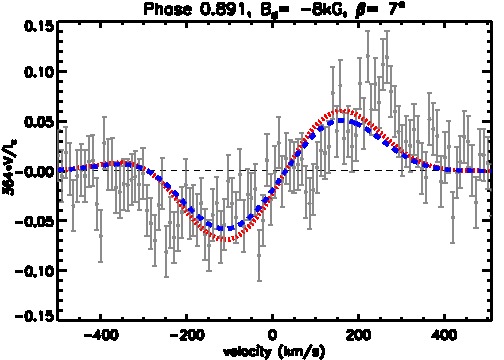}
\includegraphics[width=1.65in]{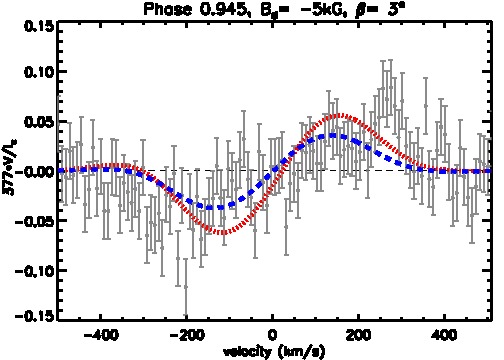}
\includegraphics[width=1.65in]{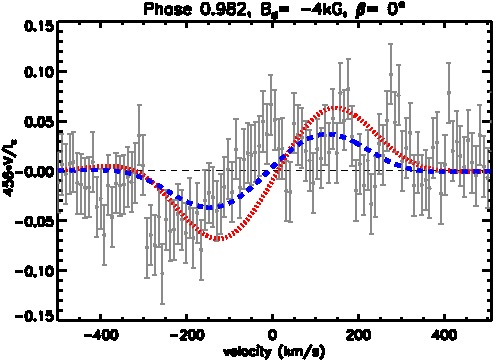}
\caption{Selected mean circularly polarised LSD Stokes $V$ profiles (grey circles) obtained from the ESPaDOnS dataset sampling the rotational cycle of HR\,5907. The error bars represent the 1$\sigma$ uncertainties for each pixel. Also shown are the individual best-fit model profiles for each phase (blue dashed) in addition to profiles corresponding to the model that provides the global maximum likelihood (red dotted) obtained from fits to the mean Stokes $V$ profiles with $B_d=10200$\,G and $\beta=7\degr$. The phase and best-fit parameters are also indicated for each observation.}
\label{lsd_comp}
\end{figure*}

Still within the framework of the ORM, our second procedure compares the observed mean LSD Stokes $V$ profiles to a large grid of synthetic profiles that are parametrised by $B_d$, $\beta$, the phase $\phi$ and $i$. The models are computed by performing a disk integration of local Stokes $V$ profiles assuming the weak field approximation and a uniform surface abundance. The parameters of the Stokes $V$ profiles were chosen to fit the mean LSD Stokes $I$ profile created from the average of all our observations. We adopted the $\phi_0$ we obtained from our longitudinal field curve fits, but find that the derived value of $\beta$ does not vary much for small changes in $\phi_0$ and that the derived values of $B_d$ only vary by a few hundred gauss. For each Stokes $V$ profile we found the parameters that provided the lowest $\chi^2$ for each observation. Using a Bayesian framework, we then combined the $\chi^2$ distributions obtained for each observation to find the maximum likelihood model. In Fig.~\ref{lsd_comp} we compare the synthetic profiles of our maximum likelihood model (dotted red) with the best-fit model for each observation (dashed blue). The quality of the fits are similar, showing that a single dipole configuration can, for most phases, roughly reproduce the observed mean profiles. However, as with our fits to the longitudinal field curve, some phases are rather poorly fit by the ORM model.

The maximum likelihood model was found with $B_d=10400^{+280}_{-350}$\,G and $\beta=7\pm1\degr$, where the uncertainties represent the 95 percentile range. While the formal uncertainties are quite low, we note that the individual best-fits to each observation have best-fit parameters ranging from $B_d\sim4000-25000$\,G and $\beta\sim0-15\degr$. 

Therefore, modelling of the longitudinal field curve and Stokes $V$ profiles indicate a strong magnetic field nearly aligned with the rotation axis. However, the observations at different phases differ quite significantly from the overall best-fit model. This will be further discussed in Sect.~\ref{disc_sec}. 

\section{The Magnetosphere}\label{magneto_sec}
Within the context of the magnetosphere model, the common picture is that the variability that is observed in hydrogen lines among Bp stars results from the circumstellar plasma. This plasma is centrifugally supported and magnetically confined into clouds that are forced into co-rotation with the host star \citep[e.g. ][]{shore90, shore+90, shore93, town05, ud08}. The circumstellar gas is seen in emission (usually in hydrogen and a few other lines) at quadrature phases, and in absorption when passing in front of the star.

\begin{figure*}
\centering
\includegraphics[width=6.7in]{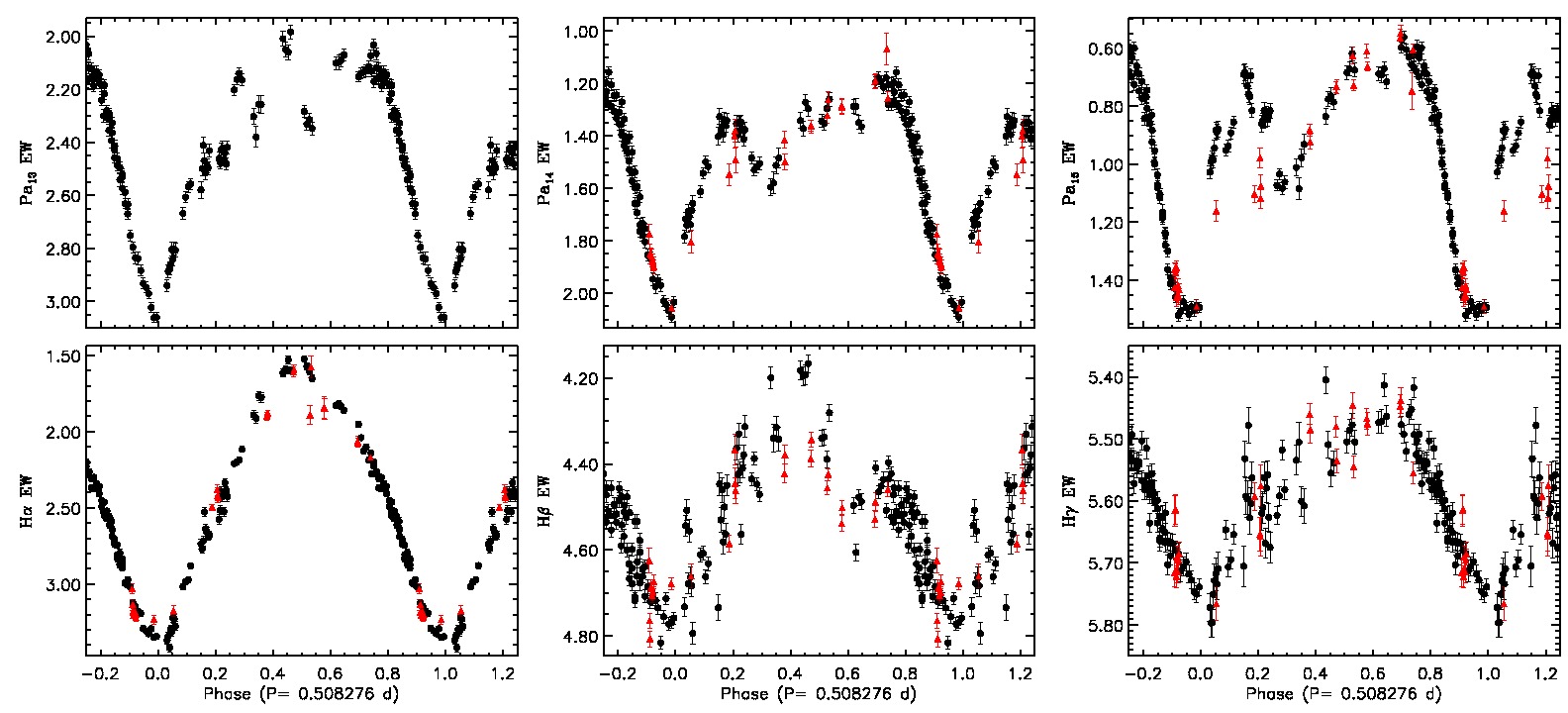}
\caption{Phased equivalent width variations of hydrogen lines.}
\label{hyd_ew}
\end{figure*}

As with the photospheric variability, we begin by first analysing the EW variations in the hydrogen Paschen and Balmer lines. As discussed in Sect.~\ref{ephem_sec}, HR\,5907 shows strong EW variations due to the circumstellar emission in H$\alpha$. In Fig.~\ref{hyd_ew} the EW variations for a number of Paschen and Balmer lines are also depicted. The strongest variation and simplest pattern is that of H$\alpha$ (Fig.~\ref{hyd_ew} bottom left panel and Fig.~\ref{phot_mag_comp} bottom panel), which shows a single emission peak at phase 0.5. This basic pattern is also present in the higher Balmer lines such as H$\beta$ and H$\gamma$ (Fig.~\ref{hyd_ew} bottom panels), but with much more noise. However, there also appears to be a second, smaller absorption dip in the H$\beta$ EW variations at phase 0.6 that is not present in the other Balmer EW variations. If we now look at the Paschen lines, which probe the magnetosphere at a different optical depth (Fig.~\ref{hyd_ew} top panels), we see a much more complex EW curve, composed of many smaller local absorption dips in addition to the single large emission peak. The overall structure of the EW curve for the Paschen lines is similar, but the amplitude of the local absorption variability varies significantly among the different lines available in the ESPaDOnS/UVES spectral range. We find the Pa$_{15}$ line to show the strongest overall variability, but Pa$_{14}$ has less scatter in the EW variations showing a local maximum emission at phase 0.18, followed by a local minimum emission at phase 0.33. Another local maximum in the emission is found at phase 0.69, but there also appears to be some additional structure to the EW curve between phases 0.33 and 0.69 that is poorly sampled by our data.

\begin{figure*}
\centering
\includegraphics[width=3.46in]{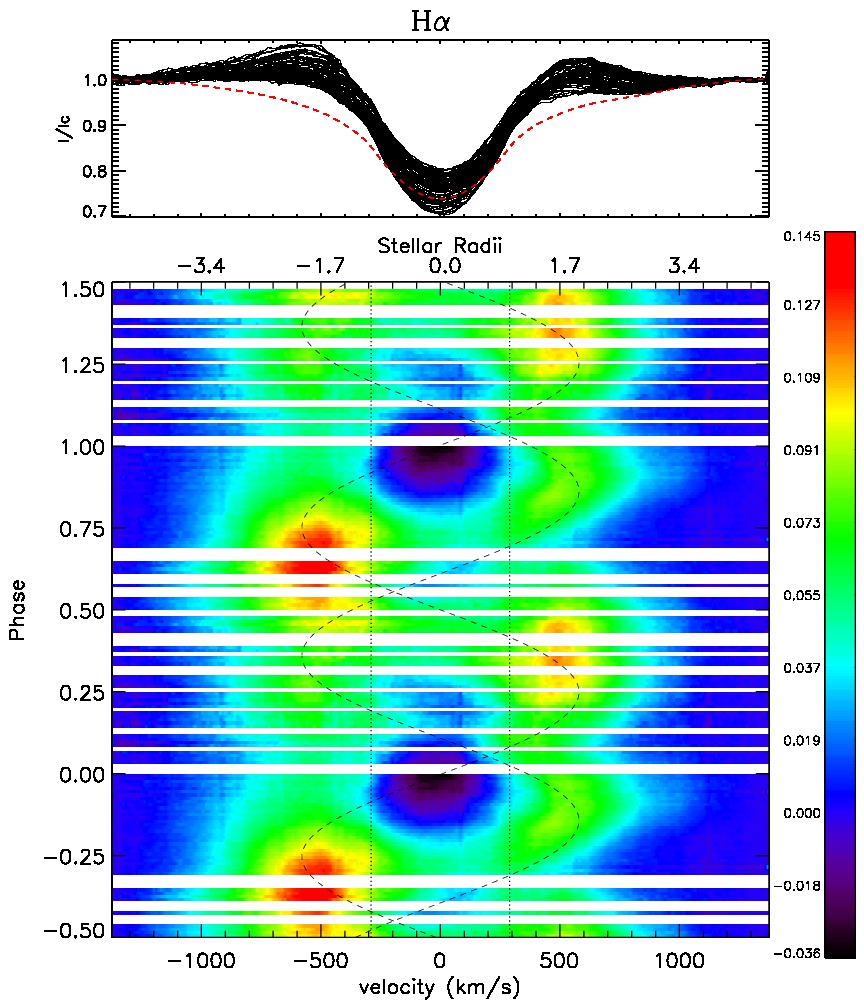}
\includegraphics[width=3.46in]{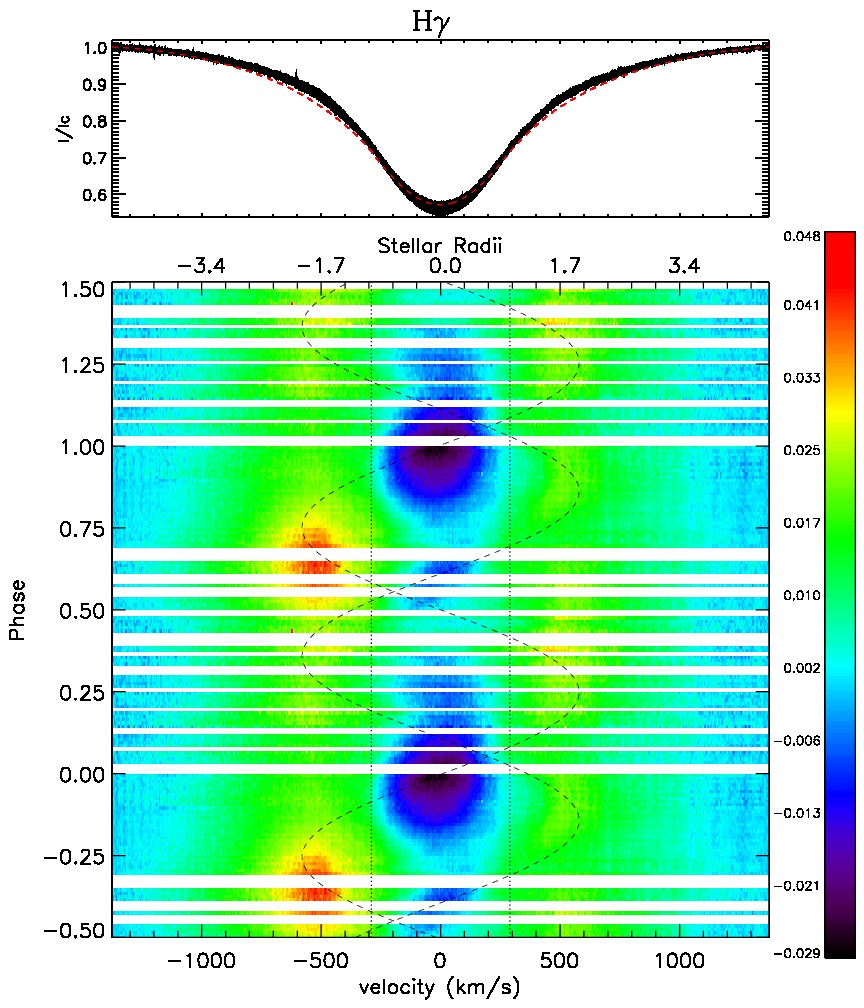} \\
\includegraphics[width=3.46in]{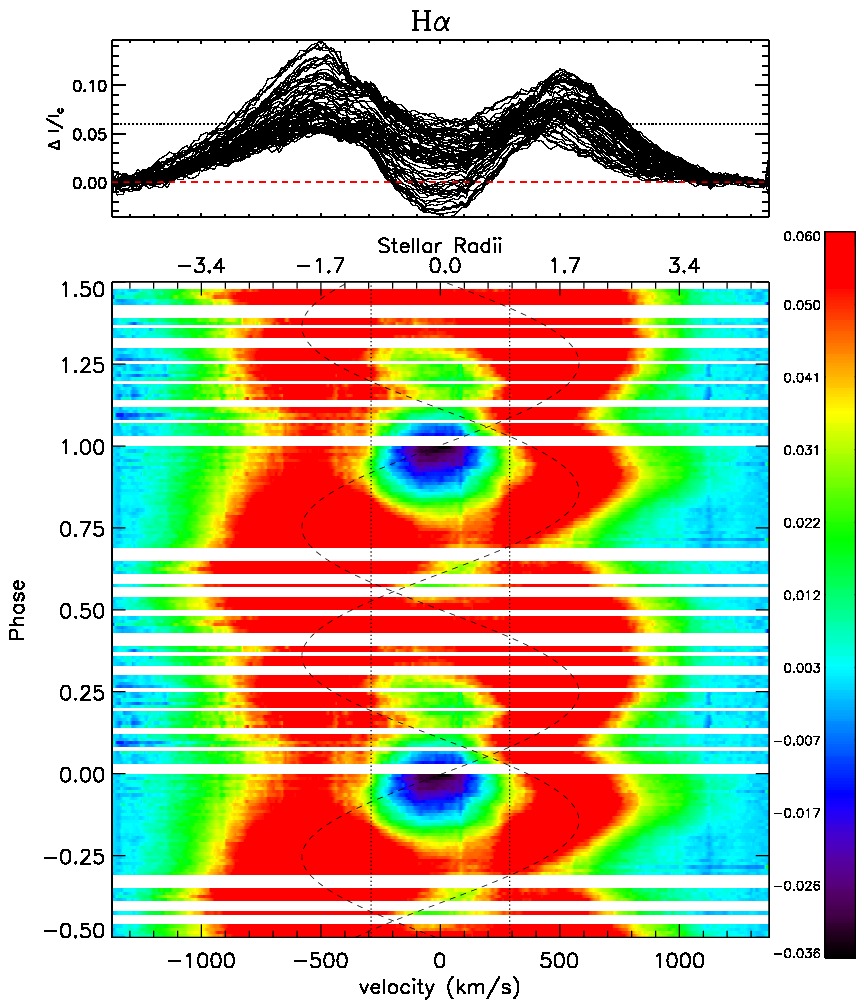}
\includegraphics[width=3.46in]{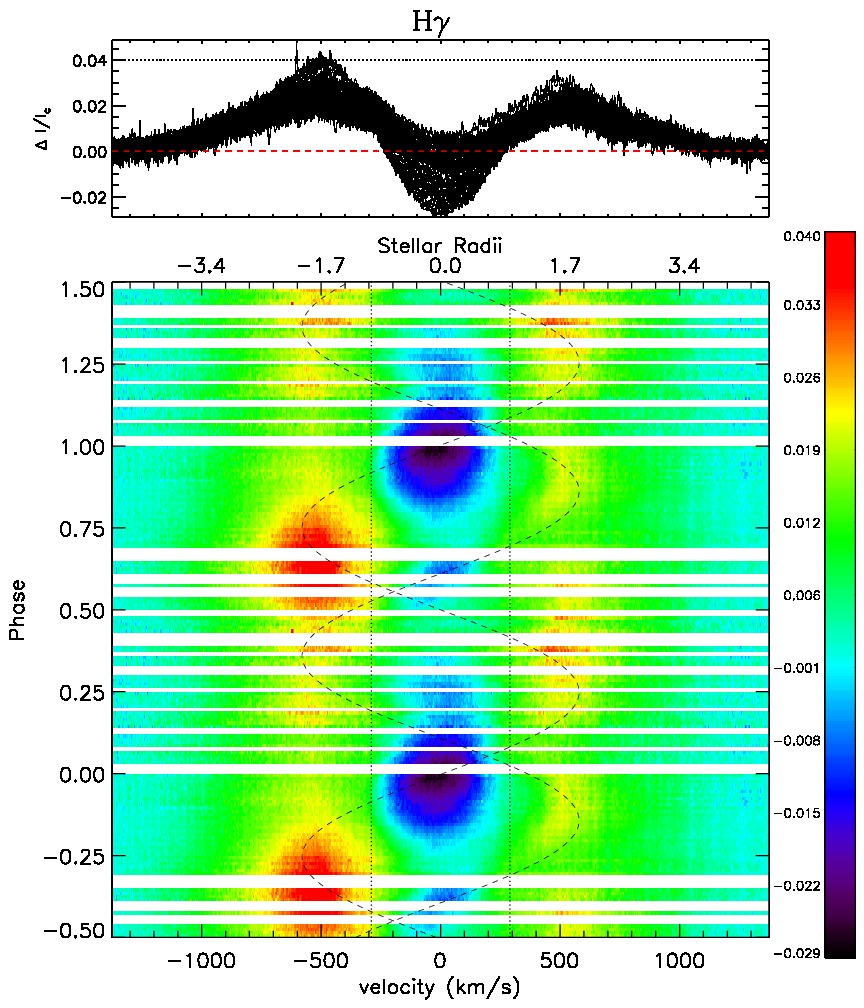}
\caption{Phased observational variations of the circumstellar magnetosphere. Shown are the variations in the indicated Balmer lines after subtracting the best-fit BK3 photospheric model (dashed red). The upper panels have a colour scheme chosen to show the full dynamic range of the emission variation, while the dynamic range in the lower panels are chosen to highlight the emission (green to red) features versus absorption (blue to black) features. Also shown are dashed vertical lines to indicate the rotational velocity (and radius) of HR\,5907 and dashed curves to highlight the motion of the emission features as discussed in the text.}
\label{balm_emis}
\end{figure*}

\begin{figure*}
\centering
\includegraphics[width=3.46in]{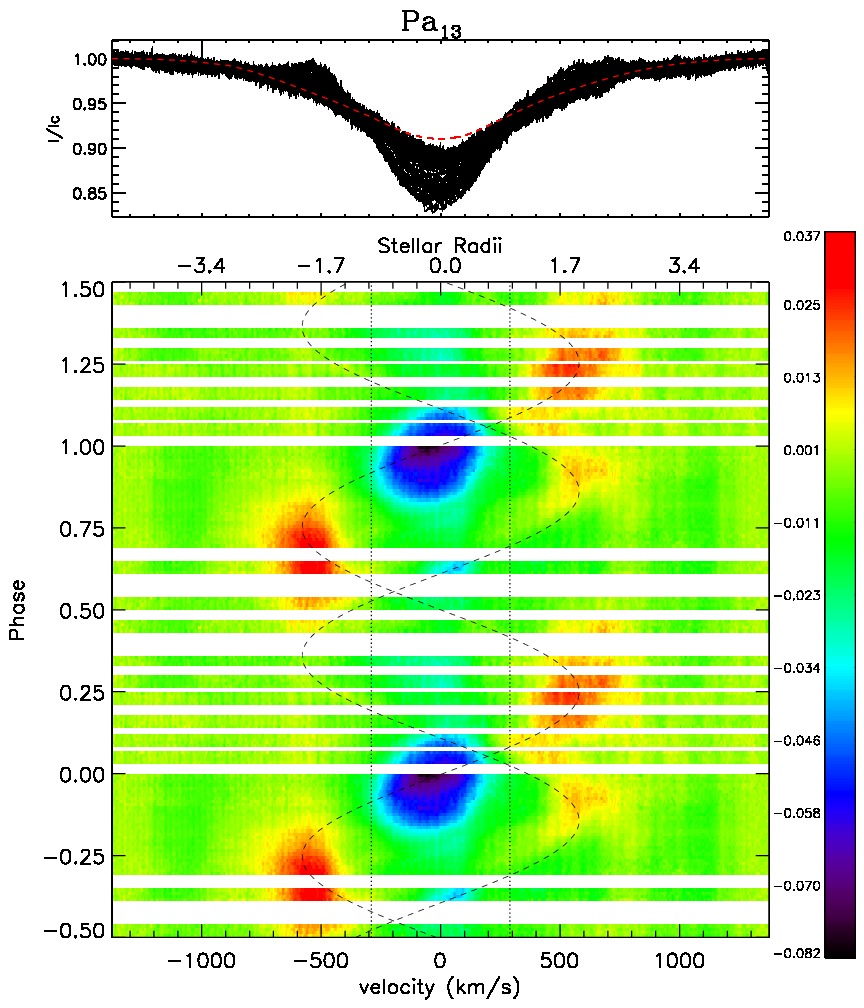}
\includegraphics[width=3.46in]{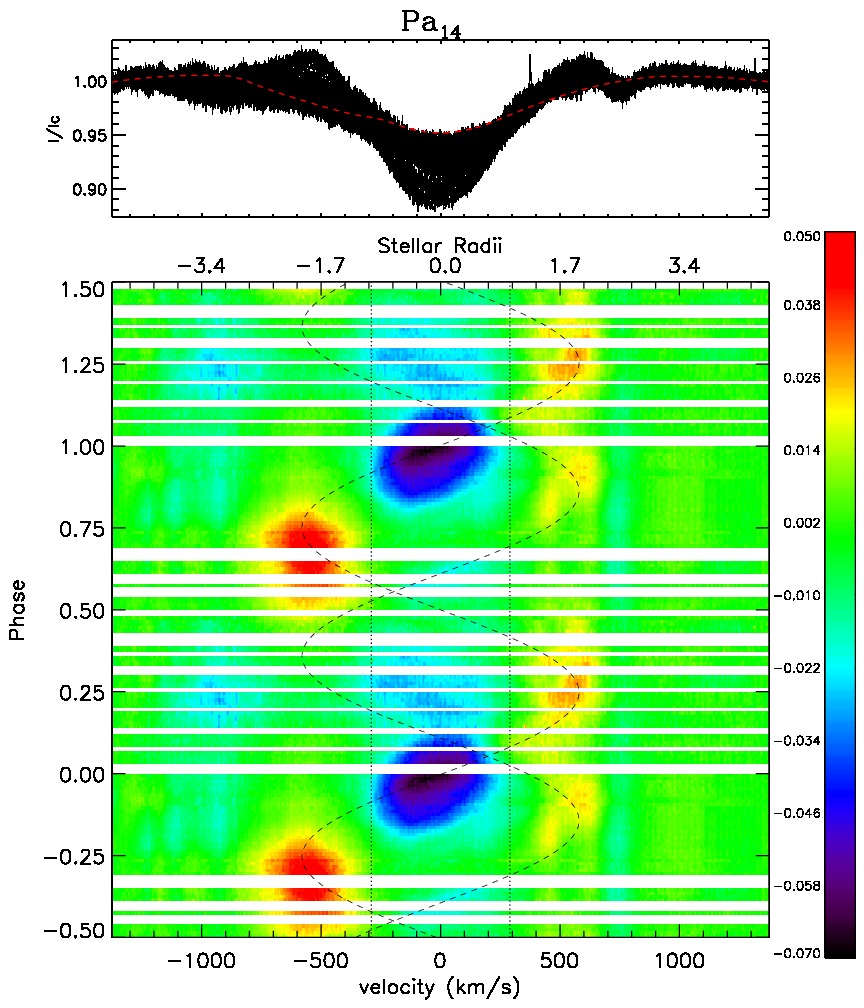}\\
\includegraphics[width=3.46in]{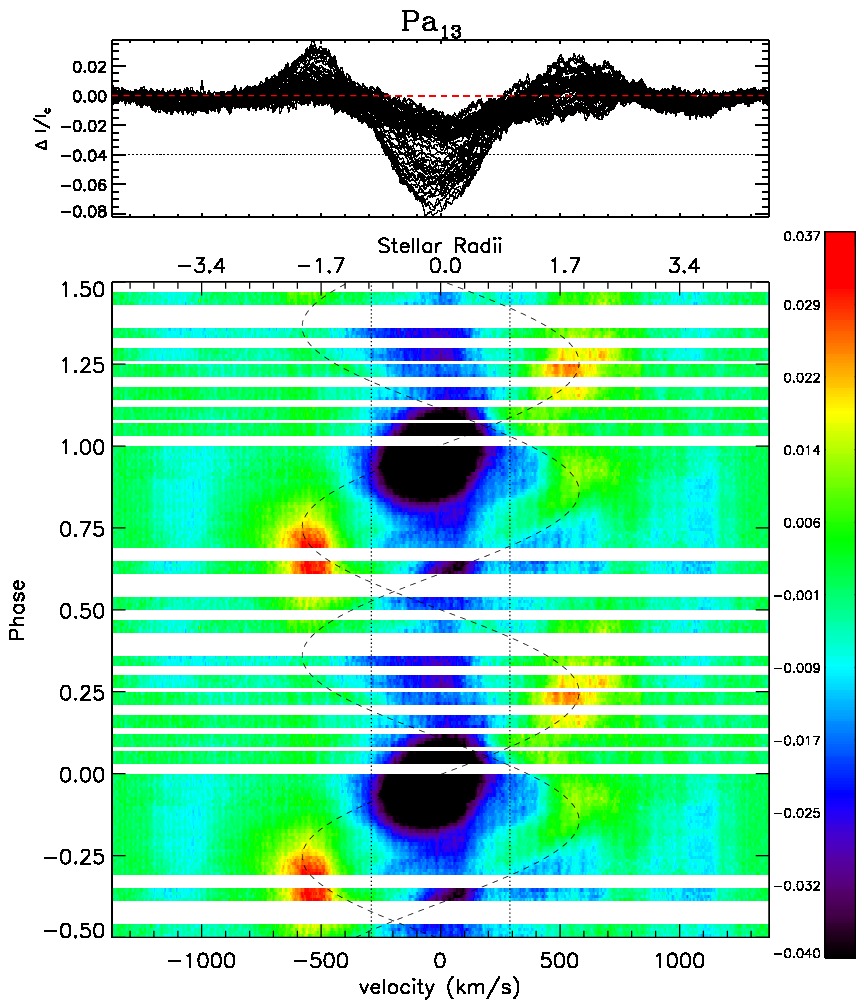}
\includegraphics[width=3.46in]{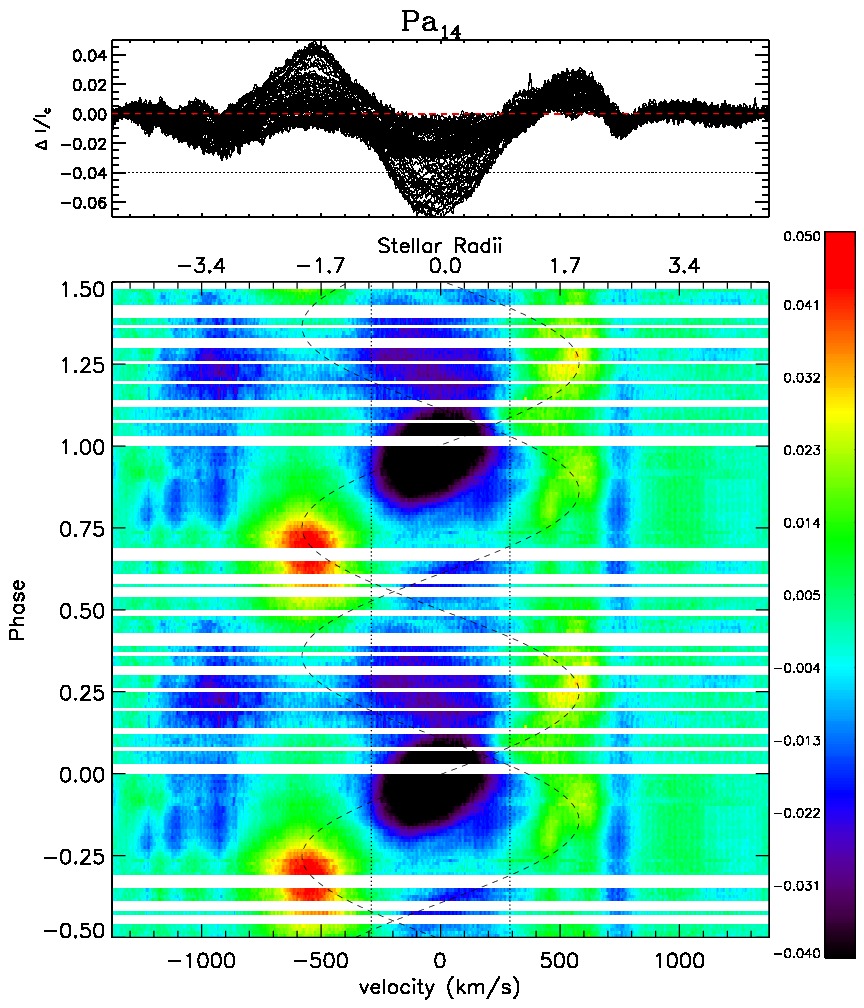}
\caption{Phased observational variations of the circumstellar magnetosphere. Shown are the variations in the indicated Paschen lines after subtracting the best-fit TLUSTY photospheric model (dashed red). The upper panels have a colour scheme chosen to show the full dynamic range of the emission variation, while the dynamic range in the lower panels are chosen to highlight the emission (green to red) features versus absorption (blue to black) features. Also shown are dashed vertical lines to indicate the rotational velocity (and radius) of HR\,5907 and dashed curves to highlight the motion of the emission features as discussed in the text.}
\label{pasch_emis}
\end{figure*}

We can further our understanding of the magnetospheric structure by studying the dynamic spectra shown in Figs.~\ref{balm_emis} and \ref{pasch_emis}. However, unlike the dynamic spectra shown in Sect.~\ref{spec_var_sec}, these dynamic spectra only include contributions from the emission, as a theoretical photospheric profile has been subtracted from each observation. In Fig.~\ref{balm_emis}, we subtracted profiles corresponding to the best-fit BK3 model determined in Sect.~\ref{fund_sect}. Since these models don't extend into the Paschen region, we found the TLUSTY model \citep{lanz07} that provided the best-fit to the Balmer lines of our BK3 model. This best-fit model ($T_{\rm eff}=19$\,kK and $\log(g)=4.25$) was found to reproduce the wings of the BK3 model Balmer lines well, but has a core that is less deep than the same BK3 model.

From the H$\alpha$ dynamic spectra we find that the main occultation occurs at phase 0.0, travelling redward. If we interpret this feature as a dense cloud in the circumstellar environment, then the fact that the occultation occupies a relatively broad velocity range during the eclipse indicates the cloud is azimuthally extended. Immediately following the broad absorption feature is a narrower feature that is still in emission in H$\alpha$ (as highlighted in the bottom panel of Fig.~\ref{balm_emis}), but is in absorption in H$\gamma$ and the Paschen lines of Fig.~\ref{pasch_emis}. This feature is travelling blueward and crosses $v_{sys}=0$\,km\,s$^{-1}$ at phase 0.21. Another occultation is clearly visible in the dynamic spectrum of H$\gamma$ moving redward and crossing $v_{sys}=0$\,km\,s$^{-1}$ at phase 0.59. Two clear emission peaks are also visible; the first feature appears at positive velocities, with the centre of the feature occurring at phase 0.36. A second blueward feature occurs at phase 0.62. We have included curves in Figs.~\ref{balm_emis} and \ref{pasch_emis} that indicate the potential orbit for a given cloud if it is in rigid rotation. The curves were selected such that they pass through the blueward-migrating occultations and reach the same velocity extrema as the brightest emission features in the Paschen lines (which approximately correspond to the Kepler radius; see Sect.~\ref{disc_sec} for further details). From these orbital curves it seems likely that these two features are the same cloud viewed at opposite sides of the star, but the expected phase difference should be 0.5 and not 0.31 as measured. The mismatch between these curves and the migrating features seen in the dynamical spectra suggests that the features do not trace single, discrete clouds in the rigidly rotating magnetosphere, but rather arise from the combined effects of multiple, optically thick clouds distributed over a range of azimuths. The observed discrepancy between the maximum emission of this cloud at opposite quadratures is also qualitatively explained if the cloud is optically thick. Another redward emission feature is also visible, reaching a maximum redward velocity at phase 0.90 in H$\alpha$ and H$\gamma$. This feature appears to be related to the blueward emission feature that is more visible in H$\gamma$ at phase 0.38. If we interpret the emission features resulting from two distinct emission clouds in the circumstellar disk, it would imply that the clouds have a phase separation of $\sim$0.6.

In the dynamic spectra of the Paschen lines (Fig.~\ref{pasch_emis}) we still see the same basic features as found in the dynamic spectra of the Balmer lines - two strong emission features, followed by a third weaker feature. However, the phases at which the emission features occur are not consistent with the phases as observed in the Balmer lines. The centre of the first redward emission feature is observed at phase 0.23, the blueward feature is found to occur at phase 0.66, and the last redward feature is found at phase 0.91. The differences in the relative phasing of the features may be a result of subtracting the TLUSTY model and not the corresponding BK3 model, but this is unexpected as the locations of these emission peaks are far in the wings where the TLUSTY models provide the best fit to the BK3 models. There also appears to be a large disagreement in the phase difference between maximum emission, which cannot be due to using a TLUSTY model, as this is a constant spectrum subtracted from all models. The phase difference between the first redward feature and the blueward feature is 0.31 for the Balmer lines, but 0.43 in the Paschen lines. The phase difference between the blueward feature and the second redward emission peak is only 0.25 from the Paschen lines, compared to 0.28 from the Balmer lines. We believe the differences between the Balmer and Paschen dynamic spectra reflect their differing optical depths; the Paschen lines are (relatively) less optically thick, and therefore come closer to tracing individual clouds in the magnetosphere.

Since we know the circumstellar material is bound in co-rotation, we can unambiguously map radial velocity onto the projected stellar surface, as indicated by the upper horizontal axis in Figs.~\ref{balm_emis} and \ref{pasch_emis}. With this information, we can now measure the furthest extent of the magnetospheric emission from the centre of the star. We find that the emission extends out to a maximum distance of 3.8\,$R_{eq}$ at phase 0.39, 4.4\,$R_{eq}$ at phase 0.59, and 4.2\,$R_{eq}$ at phase 0.93, as measured from the H$\alpha$ dynamic spectrum. The middle of both redward features is found to occur at a distance of $\sim$1.7\,$R_{eq}$, while the blueward feature is found to occur at a distance of 1.9\,$R_{eq}$. 

We also estimated the density in the circumstellar environment following the procedure of \citet{stefl03}. Using the BK3 model as a good approximation of the photospheric profiles, we estimated the Balmer decrements $D_{34}$ and $D_{54}$ during phases where there is no apparent absorption in the dynamic spectra of H$\alpha$ ($\phi=0.45,\ 0.72$), from the residuals of the observed minus theoretical equivalent widths, corrected to absolute flux using $f_\star(\rm H\alpha)/f_\star(\rm H\beta)=0.36$ and $f_\star(\rm H\gamma)/f_\star(\rm H\beta)=1.44$, as measured from the BK3 model. The Balmer decrements were then converted to densities using the theoretical computations of \citet{will88}, which were derived for an isothermal, pure hydrogen accretion disk of 10\,000\,K (the derived densities are weakly dependent on the temperature), which is optically thin in the continuum. 

From these measurements, we find that the logarithmic density ranges from 11.3 to 13.5 particles per cm$^{3}$. There is a large discrepancy between the logarithmic densities derived using $D_{34}$, which give a value of 13.5 at both phases, compared to values computed from $D_{54}$ (11.3, 11.6). We speculate that the difference may be due to the poor normalisation of the blue wing of the H$\beta$ line and that the EW measurements of this line do not span the same velocity range as the other Balmer lines, since part of the blue wing of this line is not covered over a single order in the ESPaDOnS spectra. Another possibility is that the magnetospheric densities are very near the optically thick limit, or more likely straddling this limit, above which the measured Balmer decrements become independent of the density. 

\section{Discussion \& Conclusions}\label{disc_sec}
\begin{table}
\centering
\caption{Summary of stellar, magnetic, wind, and magnetospheric properties of HR\,5907 derived in this work.}
\begin{tabular}{lr}
\hline
\multicolumn{2}{c}{\bf Stellar Parameters} \\
Stellar Mass ($M_{\odot}$) & $5.5\pm0.5$ \\
$T_{\rm eff}$ (K)             & 17\,000 $\pm$ 1000 \\
$\log(L_\star$/$L_\odot)$    & $ 2.78\pm0.14$ \\
$P_{rot}$ (days) & $0.508276^{+0.000015}_{-0.000012}$ \\
$v_{eq}\sin i$ (km\,s$^{-1}$) & $290\pm10$ \\
Inclination (\degr) & $70\pm10$ \\
Mass ($M_\odot$) & $5.5\pm0.5$\\
Equatorial Radius ($R_{\odot}$) & $3.1\pm0.2$\\
$E(B-V)$ & 0.14 \\
$\omega$ ($\Omega/\Omega_{\rm crit}$) & 0.80 \\
$v_{eq}$ (km\,s$^{-1}$) & 309 \\
Polar radius ($R_{\odot}$) & 2.72 \\
Polar temperature (K) & 18550 \\
Equatorial temperature (K) & 15920 \\
Log polar gravity (cm\,s$^{-2}$) & 4.31 \\
Log equatorial gravity (cm\,s$^{-2}$) & 4.20 \\
\multicolumn{2}{c}{\bf Magnetic Parameters}\\
$B_d$ (from $B_\ell$) (kG) & $-15.7^{+0.9}_{-0.8}$ \\
$B_d$ (from LSD Stokes $V$) (kG) & $-10.4^{+0.4}_{-0.3}$ \\
$\beta$ (\degr) & $7^{+1}_{-2}$ \\
\multicolumn{2}{c}{\bf Wind Properties}\\
$\log \dot{M}$ ($M_\odot$\,yr$^{-1}$) & -10 \\
$v_\infty$ (km\,s$^{-1}$) & 850 \\
\multicolumn{2}{c}{\bf Magnetosphere Properties}\\
$\eta_*$ & $\sim$$10^6$ \\
$R_{\rm kep}$ ($R_{eq}$) & 2.0 \\
$R_{\rm Alf}$ ($R_{eq}$) & $\sim$31\\
Min extent of H$\alpha$ Emission ($R_{eq}$) & $\sim$1.2 \\
Max extent of H$\alpha$ Emission ($R_{eq}$) & 4.4\\
Log particle density (cm$^{-3}$) & 11.3-13.5 \\
$\tau_{\rm spin}$ (Myr) & 8 \\
\hline
\end{tabular}
\label{param_tab}
\end{table}

This paper reports extensive spectroscopic and magnetic monitoring of the rapidly-rotating, early B-type star HR\,5907, based on UVES and FORS data obtained at the VLT, ESPaDOnS data from the CFHT, and MOST photometry. These observations and their analysis were undertaken within the context of the Magnetism in Massive Stars (MiMeS) Project.

We combined newly measured MOST photometry with archival Hipparcos measurements to refine the photometric period of this star to $\sim$0.508276\,d. A period search on the spectropolarimetric data confirms this period is also present in the equivalent width variations of H$\alpha$ as well as the longitudinal magnetic field measurements. If we interpret these observations within the context of the ORM with a rigidly-rotating magnetosphere, then this period represents the rotation period of the star. A comparison of the H$\alpha$ EW curve and photometric lightcurve suggests that there is $\sim$0.05 phase offset between the minima. This phase offset is not a result of a potentially incorrect period, as this would only affect the relative phasing between the Hipparcos photometry and the current epoch of observations and this offset is still present between the MOST photometry and H$\alpha$ EW curve, indicating that there is a real phase shift between the observed minima. As with the differences between the Balmer and Paschen line profile variability, the phase shift likely arises from differences in optical depths of the continuum versus the H$\alpha$ line.

\begin{figure}
\centering
\includegraphics[width=3.3in]{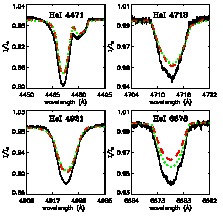}
\caption{Comparison of observed helium lines (solid black; from 24 February 2010) with our best-fit solar abundance model ($T_{\rm eff}=17$\,kK, $M_\star=5.5$\,$M_\odot$, $R_{eq}=3.1$\,$R_\odot$) computed in LTE (dashed red) and NLTE (dotted green). As the $T_{\rm eff}$ is independently constrained by the SED, this implies that HR\,5907 is a He-strong star.}
\label{he_lines_comp}
\end{figure}

Using the optical spectroscopic data, archival {\it IUE} UV observations and the estimated rotation period, we were able to constrain the stellar physical parameters by comparing the spectra to a grid of {\sc bruce/kylie} models \citep{town97}, which properly reflect the oblateness of the star due to rapid rotation. The $T_{\rm eff}$ we find ($\sim$17\,kK) is cooler than may otherwise be suggested by HR\,5907's B2.5 spectral type, but can be understood if this star is actually a He-strong star. We believe this to be the case as the observed helium line depths are deeper than the modelled helium lines computed using solar abundances (as illustrated in Fig.~\ref{he_lines_comp}). This would make HR\,5907 similar to other magnetic early B-type stars. We speculate that the helium over-abundance in HR\,5907's photosphere is likely the reason that other recent studies \citep{fre05} that use helium lines as a primary temperature indicator (even when taking into account rapid rotation) derive a much higher effective temperature ($\sim$21\,250\,K). We also investigated whether the discrepancy between the observed and modelled He lines was due to the assumption of LTE. Calculations using NLTE atmospheres (also displayed in Fig.~\ref{he_lines_comp}) are not able to resolve the discrepancy. 

As a final test, we determined the effective temperature that would be necessary to approximately fit the observed He line depths. The derived value is about 2\,kK hotter than our adopted temperature (i.e. 19\,kK). Such a temperature results in the C\,{\sc ii} 4267\,\AA\ line being about 25 percent deeper, and the UV SED being about 20 percent too large. Carbon abundances as measured from optical lines are poorly studied in Bp stars, but are typically found to be underabundant in cooler Ap stars \cite[e.g. ][]{roby90}. If $T_{\rm eff}=19$\,kK is adopted this would suggest that the carbon abundance would need to be reduced by 0.4 dex to fit the observed line depth, but this discrepancy will be somewhat mitigated if we use NLTE models, as the C\,{\sc ii} 4267\,\AA\ line is predicted to be weaker in NLTE, which would then increase the abundance necessary to fit the observed profile. However, adopting a NLTE model does not significantly affect the SED and the higher $T_{\rm eff}$ would provide a poorer fit to the UV SED. A future self-consistent analysis that takes into account rapid rotation, NLTE effects, and non-solar He abundances is necessary to remove any ambiguity.

From the low-resolution FORS and high-resolution ESPaDOnS circular polarisation measurements, we find that the mean, surface averaged, longitudinal magnetic field values are consistently negative. From the variation of these measurements we infer a dipolar magnetic field with a $\sim$16\,kG field strength at the pole and a magnetic axis nearly aligned with the rotation axis ($\beta\sim7\degr$). However, direct modelling of the mean LSD profiles of the velocity-resolved Stokes~$V$ signatures suggests a lower dipole polar field strength of $\sim$10\,kG. In any event, modelling of the circular polarisation data shows large deviations from the simple dipole model at several rotational phases. These important deviations may indicate a more complex topology to the magnetic field structure, similar to what is seen in some other magnetic B-type stars. However, the observed deviations could also be due to the strongly variable helium lines that dominate our mean LSD profiles. In fact, the lower $B_\ell$ measurements around phase 0.5 are completely consistent with what is expected if helium is less abundant at the pole and therefore more weighted towards the equatorial regions where the helium lines are stronger. A self-consistent He-abundance and magnetic field map of the stellar surface, similar to HD\,37776 \citep{koch11}, is required to unambiguously determine the magnetic field topology of HR\,5907.

Comparing the observed lightcurve with the photometric predictions of \citet{town08}, using our inferred inclination and magnetic obliquity, shows the two to be in good qualitative agreement. A lower inclination would not result in a perceivable dip in the lightcurve, unless the magnetic obliquity was about 30\degr or higher. On the other hand, if the inclination were higher, we would expect to see two dips in the lightcuve that would require a $\sim$1\,d period (which would be inconsistent with the variations observed with the $B_\ell$ measurements) as opposed to the single dip that we do observe with our $\sim$0.5\,d period. If the magnetic obliquity was much larger, we would expect to see two dips in the lightcurve, unless the inclination was lower.

As shown in Sect.~\ref{spec_var_sec}, there appear to be significant chemical abundance variations across the stellar photosphere. Helium appears to have the strongest contrast, with equivalent width variations on the order of 20 percent, which is actually much smaller than observed in other rapidly-rotating He-strong stars such as $\sigma$~Ori E or HR\,7355. Carbon, silicon, and nitrogen are also found to vary weakly. In other He-variable stars the locations of these abundance spots are correlated with the magnetic field \citep[e.g. ][]{boh88} and usually axisymetric about the magnetic equator. Therefore the weak helium variability may well reflect the observed magnetic geometry - we are nearly always viewing the same orientation of the magnetic hemispheres and would expect only small rotational modulation since the magnetic axis is nearly aligned with the rotation axis. While the absolute helium variability might be weak compared to other rapidly-rotating He-strong stars, the helium surface distribution may vary enough to cause photometric fluctuations. The small dip that is evident in the photometric lightcurve (at about phase 0.65) also coincides with a small increase in helium absorption at this phase, as shown in Fig.~\ref{he1_ew}. 

Using the stellar parameters as derived in this paper and wind parameters estimated from these stellar parameters (included in Table~\ref{param_tab} and obtained using formulae from \citet{cak75} and \citet{vink01}) , we find that the wind magnetic confinement parameter \citep{ud02} $\eta_*=B^2_{\rm eq}R^2/\dot{M}V_{\infty}\sim10^6$, which indicates that the circumstellar material is strongly confined out to large distances from the star ($R_{\rm alf}=\eta_*^{1/4}\sim30-40\,R_{eq}$). The Kepler, or co-rotation, radius \citep{ud08} $R_{\rm kep}=3/2\,\omega^{-2/3}\,R_{pole}=2.0$\,$R_{eq}$, implying that beyond this radius, the material is centrifugally supported. We find that the locations of the emission peaks in Balmer and Paschen lines agree well with this value. We also computed the predicted spin-down time using Eq. (25) of \citet{ud09}, $\tau_{\rm spin}\sim8$\,Myr, which is considerably longer than the potential age of HR\,5907. 

In summary, this study reports the detection of a large-scale, organised magnetic field with a polar surface intensity of 10-16\,kG in HR\,5907. Interpreted within the context of the ORM, we find photometric, H$\alpha$ and longitudinal magnetic field variations consistent with a 0.508276\,d rotational period, making this the shortest period, non-degenerate, magnetic massive star known to date. We find evidence of surface abundance variations in helium, carbon, silicon, and nitrogen. 

This is one of only a few known rapidly-rotating magnetic massive stars that show strong emission variations due to a magnetosphere. The only other massive star with a comparable rotation period is HR\,7355, which is believed to have a magnetic geometry more similar to $\sigma$ Ori E (a magnetic obliquity closer to 90\degr), which is very different from this star. HR\,5907 is therefore an ideal target for comparison with the predictions of rigidly rotating magnetosphere models \citep{town05, town08} and a great testbed for studying the effects of the magnetic field orientation on angular momentum loss and magnetic spin-down.

\section*{ACKNOWLEDGEMENTS}
JHG acknowledges financial support in the form of an Alexander Graham Bell Canada Graduate Scholarship from the Natural Sciences and Engineering Research Council of Canada (NSERC). GAW, JMM, AFJM, DBG and SMR acknowledge support from NSERC. RHDT acknowledges support from NSF grants AST-0904607 and AST-0908688. The authors thank Dr. Stefano Bagnulo for his guidance in the reduction of FORS data, Dr. John Landstreet for helpful discussion and Phil Landry for assistance in the period analysis. The ``Magnetism in Massive Stars" (MiMeS) project is supported by the CFHT, TBL and ESO (through the allocation of telescope time). We thank the CFHT/QSO and ESO operations staff for their efficiency at collecting data for this challenging target.

\label{lastpage}

\begin{thebibliography}{}
\bibitem[Bagnulo et al.(2002)]{bagn02}Bagnulo S. et al., 2002, A\&A, 389, 191
\bibitem[Bagnulo et al.(2009)]{bagn09}Bagnulo S., Landolfi M., Landstreet J.D., Degl'Innocenti E., Fossati L., Sterzik M., 2009, PASP, 121, 993
\bibitem[Bertelli et al.(2009)]{bert09}Bertelli G., Nasi E., Girardi L., Marigo P., 2009, A\&A, 508, 355
\bibitem[Bohlender et al.(1987)]{boh87}Bohlender D.A., Landstreet J.D., Brown D.N., Thompson I.B., 1987, ApJ, 323, 325
\bibitem[Bohlender \& Landstreet(1988)]{boh88}Bohlender D.A., Landstreet J.D., 1988, in G. Cayrel de Strobel, M. Spite (eds), Proc. IAUS Symp. 132. The Impact of Very High S/N Spectroscopy on Stellar Physics, Kluwer, Dordrecht, p. 309
\bibitem[Borra \& Landstreet(1973)]{bor73}Borra E.F. Landstreet J.D., 1973, ApJ, 185, L139
\bibitem[Castor, Abbott \& Klein(1975)]{cak75}Castor J.I., Abbott D.C., Klein R.I., 1975, ApJ, 195, 157
\bibitem[Cardelli, Clayton \& Mathis(1989)]{card89}Cardelli J.A., Clayton G.C., Mathis J.S., 1989, ApJ, 345, 245
\bibitem[Donati et al.(1997)]{don97}Donati J.-F. et al., 1997, MNRAS, 291, 658
\bibitem[Donati et al.(2002)]{don02}Donati J.-F. et al., 2002, MNRAS, 333, 55
\bibitem[Donati et al.(2006)]{don06}Donati J.-F. et al., 2006, MNRAS 365, L6
\bibitem[Fr{\' e}mat et al.(2005)]{fre05}Fr\'{e}mat Y., Zorec J., Hubert A.-M., Floquet M., 2005, A\&A, 305, 320
\bibitem[Harmanec(1998)]{harm98}Harmanec P., 1998, A\&A, 335, 173
\bibitem[Hern{\' a}ndez et al.(2005)]{hern05}Hern{\' a}ndez J., Calver N., Hartmann L., Brice{\~ n}o C., Sicilia-Aguilar A., Berlind P., 2005, ApJ, 129, 856
\bibitem[Hoffleit \& Warren(1991)]{hof91}Hoffleit D., Warren W. H. Jr., 1991, The Bright Star Catalogue, 5th Revised Ed. (New Haven, CT: Yale Univ. Observatory)
\bibitem[Hubert \& Floquet(1998)]{hub98}Hubert A.M., Floquet M., 1998, A\&A, 335, 565
\bibitem[Hynes(2011)]{hynes}Hynes R., 2011, \url{http://www.astro.soton.ac.uk/~rih/applets/MagCalc.html}
\bibitem[Jaschek \& Egret(1982)]{jasch82}Jaschek M., Egret D., ViZieR Online Data Catalogue
\bibitem[Jehnin \& O`Brien(2008)]{jehn08}Jehnin E., O`Brien K., 2008, FORS User Manual (VLT-MAN-ESO-13100-1543), ESO, Garching, edition for period 81
\bibitem[Kharchenko \& Roeser(2009)]{khar09}Kharchenko N.V., Roeser S., 2009, ViZieR Online Data Catalogue
\bibitem[Kochukhov et al.(2011)]{koch11}Kochukhov O., Lundin A., Romanyuk I., Kudryatisev D., 2011, ApJ, 726, 24
\bibitem[Kurucz(1992)]{kur92}Kurucz R.L., 1992, in B. Barbuy \& A Renzini ed., The Stellar Populations of Galaxies Vol. 149 of IAU Symposium, Model Atmospheres for Population Synthesis. p. 225
\bibitem[Landstreet \& Borra(1978)]{land78}Landstreet J.D., Borra E.F., 1978, ApJ, 224, L5
\bibitem[Lanz \& Hubeny(2007)]{lanz07}Lanz T., Hubeny I., 2007, ApJS, 169, 83
\bibitem[Lef{\`e}vre et al.(2009)]{lef09}Lef{\`e}vre L., Marchenko S.V., Moffatt A.F.J., Acker A., 2009, A\&A, 507, 1141
\bibitem[Leone \& Umana(1993)]{leon93}Leone F., Umana G., 1993, A\&A, 268, 667
\bibitem[Lewis et al.(2009)]{lew09}Lewis N.K., Cook T.A., Wilton K.P., France K., Gordon K.D., 2009, ApJ, 706, 306
\bibitem[Maeder \& Meynet(2010)]{maeder10}Maeder A., Meynet G., 2010, NewAR, 54, 32
\bibitem[Oksala et al.(2010)]{oks10}Oksala M.E., Wade G.A., Marcolino W.L.F., Grunhut J., Bohlender D., Manset N., Townsend R.H.D., 2010, MNRAS, 405, L510
\bibitem[Papaj, Wegner \& Krelowski(1991)]{papaj91}Papaj J., Wegner W., Krelowski J., 1991, MNRAS, 252, 403
\bibitem[Pedersen \& Thomsen(1977)]{ped77}Pedersen H., Thomsen B., 1977, A\&AS, 30, 11
\bibitem[Press et al.(1992)]{press92}Press W.H., Teukolsky S.A., Vetterling W.T., Flannery B. P., 1992, Numerical Recipes in FORTRAN, 2nd edn. Cambridge University Press, Cambridge
\bibitem[Reiners et al.(2000)]{rein00}Reiners A., Stahl O., Wolf B., Kaufer A., Rivinius T., 2000, A\&A, 363 585
\bibitem[Rivinius et al.(2010)]{rivi10}Rivinius T., Szeifert T., Barrera L., Townsend R.H.D., {\v S}tefl S., Baade D., 2010, MNRAS, 405 L46
\bibitem[Roby \& Lambert(1990)]{roby90}Roby S.W., Lambert D.L., 1990, ApJS, 73, 67
\bibitem[Schaller et al.(1992)]{schal92}Schaller G., Schaerer D., Meynet G., Maeder A., 1992, A\&AS, 96, 269
\bibitem[Schwarzenberg-Czerny(1996)]{sc96}Schwarzenberg-Czerny A., 1996, ApJL, 460, 107
\bibitem[Shore et al.(1990)]{shore+90}Shore S.N. Brown D.N., Sonneborn G., Landstreet J.D., Bohlender D.A., 1990, ApJ, 348, 242
\bibitem[Shore \& Brown(1990)]{shore90}Shore S.N., Brown D.N., 1990, ApJ, 365, 665
\bibitem[Shore(1993)]{shore93}Shore S.N., 1993, in Dworetsky M.M. Castelli F., Faraggiana R., eds, ASP Conf. Ser. Vol. 44, IAU Coll. 138, Peculiar Versus Normal Phenomena in A-type and Related Stars. Astron. Soc. Pac., San Francisco, p. 528
\bibitem[Silvester et al.(2009)]{silv09}Silvester J. et al., 2009, MNRAS, 398, 1505
\bibitem[Smith \& Bohlender(2007)]{smith07}Smith M.A., Bohlender D.A., 2007, A\&A, 475, 1027
\bibitem[Townsend(1997)]{town97}Townsend R.H.D., 1997, MNRAS, 284, 839
\bibitem[Townsend \& Owocki(2005)]{town05}Townsend R.H.D., Owocki S.P., 2005, MNRAS, 357, 251
\bibitem[Townsend(2008)]{town08}Townsend R.H.D., 2008, MNRAS, 389, 559
\bibitem[ud-Doula \& Owocki(2002)]{ud02}ud-Doula A., Owocki S.P., 2002, ApJ, 576, 413
\bibitem[ud-Doula, Owocki \& Townsend(2008)]{ud08}ud-Doula A., Owocki, S.P., Townsend, R.H.D., 2008, MNRAS, 385 97
\bibitem[ud-Doula, Owocki \& Townsend(2009)]{ud09}ud-Doula A., Owocki S.P., Townsend R.H.D., 2009, MNRAS, 392, 1022
\bibitem[van Leeuwen(2007)]{vanle07}van Leeuwen F., 2007, A\&A, 474, 653
\bibitem[Vink, de Koter \& Lamers(2001)]{vink01}Vink J.S., de Koter A., Lamers H.J.G.L.M., 2001, A\&A, 369, 574
\bibitem[{\v S}tefl et al.(2003)]{stefl03}{\v S}tefl S., Baade D., Rivinius T., Otero S., Stahl O., Budovi{\v c}ov{\' a} A., Kaufer A., Maintz M., 2003, A\&A, 402, 253
\bibitem[Wade et al.(2011)]{wade11}Wade G.A. and the MiMeS Collaboration, 2011, proceedings of ``Magnetic Stars", Special Astrophysical Observatory, in press (arXiv :1012.2925)
\bibitem[Walborn(1982)]{walb82}Walborn N.R., 1982, PASP, 94, 322
\bibitem[Williams \& Shipman(1988)]{will88}Williams G.A., Shipman H.L., 1988, ApJ, 326, 738
\end{thebibliography}
\end{document}